\documentclass[fleqn,usenatbib]{mnras}

\usepackage{newtxtext,newtxmath}

\usepackage[T1]{fontenc}
\usepackage{ae,aecompl}

\usepackage{graphicx}	
\usepackage{amsmath}	
\usepackage{amssymb}	
\usepackage{pdflscape}

\title[Timing of 21 LOTAAS pulsars]{The LOFAR Tied-Array All-Sky Survey: Timing of 21 pulsars including the first binary pulsar discovered with LOFAR}

\author[C. M. Tan et al.]{C. M. Tan,$^{1,2,6,7}$\thanks{E-mail: chia.tan@mcgill.ca}
C. G. Bassa,$^{3}$
S. Cooper,$^{1}$
J. W. T. Hessels,$^{4,3}$
V. I. Kondratiev,$^{3,5}$
\newauthor
D. Michilli,$^{4,3,6,7}$
S. Sanidas,$^{1}$
B. W. Stappers,$^{1}$
J. van Leeuwen,$^{3,4}$
J. Y. Donner,$^{8,9}$
\newauthor
J.-M. Grie{\ss}meier,$^{10,11}$
M. Kramer,$^{9,1}$
C. Tiburzi,$^{3}$
P. Weltevrede,$^{1}$
B. Ciardi,$^{11}$
\newauthor
M. Hoeft,$^{12}$
G. Mann,$^{13}$
A. Miskolczi,$^{14}$
D. J. Schwarz,$^{8}$
C. Vocks,$^{13}$
O. Wucknitz$^{9}$
\\
$^{1}$Jodrell Bank Centre for Astrophysics, University of Manchester, Oxford Road, Manchester, M13 9PL, United Kingdom\\
$^{2}$SKA Organisation, Jodrell Bank Observatory, Lower Withington, Macclesfield, Cheshire, SK11 9FT, UK\\
$^{3}$ASTRON, Netherlands Institute for Radio Astronomy, Oude Hoogeveensedijk 4, 7991 PD, Dwingeloo, The Netherlands\\
$^{4}$Anton Pannekoek Institute for Astronomy, University of Amsterdam, Science Park 904, 1098 XH Amsterdam, The Netherlands\\
$^{5}$Astro Space Centre, Lebedev Physical Institute, Russian Academy of Sciences, Profsoyuznaya Str. 84/32, Moscow 117997, Russia\\
$^{6}$Department of Physics, McGill University, 3600 University Street, Montr\'{e}al, QC H3A 2T8, Canada\\
$^{7}$McGill Space Institute, McGill University, 3550 University Street, Montr\'{e}al, QC H3A 2A7, Canada\\
$^{8}$Fakult\"{a}t f\"{u}r Physik, Universit\"{a}t Bielefeld, Postfach 100131, 33501 Bielefeld, Germany\\
$^{9}$Max-Planck-Institut f\"{u}r Radioastronomie, Auf dem H\"{u}gel 69, D-53121 Bonn, Germany\\
$^{10}$LPC2E - Universit\'{e} d'Orl\'{e}ans /  CNRS, 45071 Orl\'{e}ans cedex 2, France\\
$^{11}$Station de Radioastronomie de Nan\c{c}ay, Observatoire de Paris, PSL Research University, CNRS, Univ. Orl\'{e}ans, OSUC, 18330 \\
Nan\c{c}ay, France\\
$^{12}$Max-Planck-Institut f\"{u}r Astrophysik, Postfach 1317, D-85741 Garching, Germany\\
$^{13}$Th\"{u}ringer Landessternwarte, Sternwarte 5, 07778 Tautenburg, Germany\\
$^{14}$Leibniz-Institut f\"{u}r Astrophysik Potsdam, An der Sternwarte 16, 14482 Potsdam, Germany\\
$^{15}$Ruhr-University Bochum, Faculty of Physics and Astronomy, Astronomical Institute, 44780 Bochum, Germany\\
}

\date{Accepted XXX. Received YYY; in original form ZZZ}

\pubyear{2018}

\begin{document}
\label{firstpage}
\pagerange{\pageref{firstpage}--\pageref{lastpage}}
\maketitle

\begin{abstract}
We report on the multi-frequency timing observations of 21 pulsars discovered in the LOFAR Tied-Array All-Sky Survey (LOTAAS). The timing data were taken at central frequencies of 149\,MHz (LOFAR) as well as 334 and 1532\,MHz (Lovell Telecope). The sample of pulsars includes 20 isolated pulsars and the first binary pulsar discovered by the survey, PSR\,J1658$+$3630. We modelled the timing properties of the pulsars, which showed that they have, on average, larger characteristic ages. We present the pulse profiles of the pulsars across the three observing bands, where PSR\,J1643$+$1338 showed profile evolution that appears not to be well-described by the radius-to-frequency-mapping model. Furthermore, we modelled the spectra of the pulsars across the same observing bands, using a simple power law, and found an average spectral index of $-1.9 \pm 0.5$. Amongst the pulsars studied here, PSR\,J1657$+$3304 showed large flux density variations of a factor of 10 over 300 days, as well as mode changing and nulling on timescales of a few minutes. We modelled the rotational and orbital properties of PSR\,J1658$+$3630, which has a spin period of 33\,ms in a binary orbit of 3.0\,days with a companion of minimum mass of 0.87$M_{\odot}$, likely a Carbon-Oxygen or Oxygen-Neon-Magnesium type white dwarf. PSR\,J1658$+$3630 has a dispersion measure of 3.0 pc cm$^{-3}$, making it possibly one of the closest binary pulsars known. 
\end{abstract}




\section{Introduction}

Since the first discovery of radio pulsars more than 50 years ago~\citep{hbp+68}, extensive studies have been done to understand their properties. These include measuring their rotational, spectral and emission properties across a large range of radio frequencies. While the first pulsar was discovered at a low radio frequency of 81\,MHz, pulsar science has so far been predominantly conducted at radio frequencies above 300\,MHz. This is primarily due to the the effects of dispersion and scattering of pulsar signals by the interstellar medium (ISM) being more pronounced at low frequencies. In regions of the Galaxy with dense ISM, the pulsar signal will be smeared out to the point of being undetectable as pulses.

Recently, there has been a resurgence in pulsar observations at low frequencies below 300\,MHz, driven by the new generation of low-frequency radio telescopes such as the LOw Frequency ARray~\citep[LOFAR;][]{hwg+13}, the Murchison Widefield Array~\citep[MWA;][]{tgb+13} and the Long-Wavelength Array~\citep[LWA;][]{tek+12}. The improvement in computing capabilities also results in minimising the effects of dispersion on the pulsar signal across a large bandwidth, allowing for wideband study of pulsars at very low radio frequencies. 

The radio spectra of most pulsars can be described by a simple power law with a spectral index $\alpha$, where the radio flux density $S$ is related to the radio frequency $\nu$ by $S \propto \nu^{\alpha}$. Generally, pulsars have steep radio spectra ($\alpha < -1$), which makes them ideal radio sources to study at very low frequencies.~\citet{mkkw00} found a mean spectral index of $\bar{\alpha}=-1.8 \pm 0.2$ by modelling the radio spectra of 281 pulsars across a large frequency range of 0.3$-$20\,GHz.~\citet{blv13} modelled the distribution of spectral indices of radio pulsars across a frequency range of 0.4$-$6.7\,GHz and found a shallower mean spectral index of $\bar{\alpha}=-1.4$, but with a large standard deviation of unity.~\citet{jsk+18} modelled the spectra of 441 pulsars across a frequency range of 0.7$-$3.1\,GHz and found a mean spectral index of $\bar{\alpha}=-1.6$ and a standard deviation of 0.54.

Several large-scale studies of the known pulsar population with LOFAR have been conducted over recent years.~\citet{nsk+15} analysed the polarisation profiles of 20 pulsars using the LOFAR High Band Antennas (HBAs) at frequencies between 105$-$197\,MHz and compared them to observations at higher frequencies. The study highlighted the importance of low-frequency study of polarisation properties to probe both the effects of pulsar magnetosphere and the interstellar medium that are more pronounced.~\citet{phs+16} studied the pulse profiles of 100 pulsars with the LOFAR HBAs at frequencies of 120$-$167\,MHz and 26 pulsars with the LOFAR Low Band Antennas (LBAs) at frequencies of 15$-$62\,MHz, and compared them with archival observations at 350 and 1400\,MHz. They found that the frequency evolution of most of the pulsars follow the radius-frequency-mapping model~\citep[RFM model,][]{rs75,cor78}, where the width of the pulse profile of a pulsar increases with decreasing observing frequency.

\citet{bkk+16} undertook a census of 194 known, isolated pulsars off the Galactic plane (|b|>$3\deg$) using the LOFAR HBAs at frequencies of 110$-$188\,MHz, providing dispersion measure and flux density measurements, as well as flux-calibrated integrated pulse profiles across the observing bandwidth. By combining the obtained results with literature values at higher radio frequencies, they modelled the spectra of 165 pulsars and computed a mean spectral index of $\bar{\alpha} = -1.4$ (no quoted sigma), lower than the value found by~\citet{mkkw00} and~\citet{jsk+18}. This is attributed to a potential turnover or flattening of spectra at low radio frequencies. A companion study was conducted by~\citet{kvh+16}, targeting 75 millisecond pulsars (MSPs) with the LOFAR HBAs at 110$-$188\,MHz, with 9 of them observed with the LOFAR LBAs as well at 38$-$77\,MHz. They presented the pulse profiles and flux densities, as well as measured the dispersion measure (DM) of the pulsars.

Since December 2012, the LOFAR Tied-Array All-sky Survey~\citep[LOTAAS;][]{clh+14,scb+19} has been observing the whole Northern sky to search for pulsars and fast transients using the HBAs at a central observing frequency of 135\,MHz, with a bandwidth of 32\,MHz. The survey has discovered a number of pulsars with steep spectra that are otherwise difficult to detect at higher observing frequencies. As of April 2019, 1919 of 1953 planned pointings of LOTAAS were completed; this has thus far resulted in the discovery of 73 pulsars and rotating radio transients~\citep[RRATs;][]{mll+06}. As part of the effort to study the pulsar population detected by LOTAAS, timing observations have been conducted on the discoveries using both the LOFAR HBAs at 149\,MHz and the Lovell Telescope at 334 and 1532\,MHz, respectively. These observations will allow us to study the various properties of the pulsars and RRATs discovered at low frequencies across a large range of radio frequencies.

Previously,~\citet{mbc+19} reported the properties of 19 pulsars discovered by LOTAAS. They found that the sample has, on average, longer spin periods and smaller spin period derivatives, hinting that the LOTAAS pulsar discoveries are older than the known pulsar population. They also reported that the radio spectra of these pulsars are, on average, steeper, potentially linking the age of pulsars to their emission properties.

Here we report the multi-frequency analysis of 21 pulsars discovered by LOTAAS. We modelled the rotational and spectral properties, as well as studied the frequency evolution of the pulse profiles. These pulsars were first detected between 2016 November and 2017 September, and subsequent timing observations up until 2018 November have provided an adequate timeline over which to model their properties. The sample of pulsars includes PSR\,J1658$+$3630, the first binary pulsar discovered by the survey. PSRs\,J0421$+$3255, J1638$+$4005, J1643$+$1338 and J1657$+$3304 were first reported by~\citet{ttol16,ttk+17,ttm18} and these pulsars were blindly detected by LOTAAS around the same time. As of September, 2019 there are no published timing models for these pulsars. Hence, the analysis of these pulsars is presented here.

Section~\ref{sec:observations} describes the observational setup and data analysis procedure of both LOFAR and the Lovell Telescope. Section~\ref{sec:timingprop} presents the rotational properties of the pulsars derived from the timing observations. Section~\ref{sec:20pulseprofile} describes the pulse profiles of the pulsars across multiple different radio frequencies. Section~\ref{sec:spectra} presents the flux densities of the pulsars at several different radio frequencies and the modelled spectral indices. Section~\ref{sec:interestingpulsars} describes the results of the additional analysis done on two of the more interesting sources in this sample of 21 pulsars, PSR\,J1657$+$3304 and the binary pulsar PSR\,J1658$+$3630.

\section{Observations and data analysis} \label{sec:observations}

\subsection{Timing of isolated pulsars} \label{sec:timing}

Here we describe the observational strategy and setup, as well as the timing analysis of the isolated pulsars discovered by LOTAAS. Whenever a pulsar is detected and confirmed by a follow-up observation with LOTAAS (see~\citealt{scb+19} for a description of the follow-up observation strategy), it is added to the monthly LOTAAS timing campaign. In the campaign, the pulsars were observed with all 24 HBA stations of the LOFAR core. The lengths of individual observations of each pulsar depend on the brightness of the pulsar and are listed in Table~\ref{tab:observations}. The dual-polarization complex voltage data were recorded 400 sub-bands of 195.3\,kHz each, centred at an observing frequency of 149.8\,MHz. The data were then processed with the automated Pulsar Pipeline~\citep[PulP;][]{sha+11,kvh+16}, which coherently dedisperse and fold the data using \textsc{dspsr}~\citep{sb10} and \textsc{psrchive}~\citep{hsm04}, respectively to produce an \textit{archive} from each observation. The \textit{archive} files are data cubes folded with 1024 phase bins across one pulse period, five seconds time sub-integrations and 400 sub-bands across the observing band of 78.1\,MHz. The setup is more sensitive than the LOTAAS survey setup, where only six HBA stations of the LOFAR Superterp are used. 

\begin{table*}
\centering
\caption{The observing span, number of observations and observation length of each of the 20 pulsars discussed in Section~\ref{sec:timing}. PSR (disc.) denotes the name of the pulsars given by~\citet{scb+19}. The dashes indicate that the pulsars are not detected by the Lovell Telescope.}
\label{tab:observations}
\begin{tabular}{lllllll}
\hline
PSR & PSR & Span & \multicolumn{2}{l}{No. of Observations} & \multicolumn{2}{l}{Observation Length}\\
 & & & LOFAR & Lovell & LOFAR & Lovell\\
 & (disc.) & (months) & & & (minutes) & (minutes)\\
\hline
J0100$+$8023  & J0100$+$80 &	17 & 20 & -- & 15 & --\\
J0107$+$1322  & J0107$+$13 &	13 & 17 & 2$^{a}$ & 15 & 42\\
J0210$+$5845  & J0210$+$58 &	10 & 16 & -- & 10 & --\\
J0421$+$3255  & J0421$+$32 &	21 & 27 & -- & 10 & --\\
J0454$+$4529  & J0454$+$45 &	13 & 18 & -- & 10 & --\\
J1017$+$3011  & J1017$+$30 &	11 & 15 & -- & 10/15 & --\\
J1624$+$5850  & J1623$+$58 &	20 & 25 & -- & 10 & --\\
J1638$+$4005  & J1638$+$40 &	15 & 20 & -- & 10 & --\\
J1643$+$1338  & J1643$+$13 &	18 & 20 &  5 & 10 & 42/15\\
J1656$+$6203  & J1655$+$62 &	20 & 27 & -- & 10 & --\\
J1657$+$3304  & J1657$+$33 &	15 & 20 & -- & 10 & --\\
J1713$+$7810  & J1713$+$78 &	18 & 22 & -- & 10 & --\\
J1741$+$3855  & J1741$+$38 &	15 & 20 & -- & 15/10 & --\\
J1745$+$1252  & J1745$+$12 &	16 & 23 & -- & 10 & --\\
J1749$+$5952  & J1749$+$59 &	22 & 18 & 22 & 10 & 12\\
J1810$+$0705  & J1810$+$07 &	22 & 19 & 20 & 10 & 12\\
J1916$+$3224  & J1916$+$32 &	22 & 19 & 19 & 10 & 12 \\
J1957$-$0002  & J1957$-$00 &	20 & 24 & -- & 10 & --\\
J2036$+$6646  & J2036$+$66 &	11 & 15 & -- & 10 & --\\
J2122$+$2426  & J2122$+$24 &	17 & 17 & 15 & 10 & 30\\
\hline
\multicolumn{7}{l}{$^a$ These observations are only used to model the profile of the pulsar.}\\
\end{tabular}
\end{table*}

The pulsars were also all observed with the Lovell Telescope at Jodrell Bank, at a central observing frequency of 1532\,MHz. Initially, each pulsar was observed 4$-$5 times over a span of about 10 days. Each observation was between 30 min and 1 hour long. If the pulsar was detected, a lower cadence observing campaign was continued, with each source typically observed once every two weeks, depending on the availability of the telescope. The observation length is shown in Table~\ref{tab:observations}, and was chosen based on the brightness in the initial detection. The observations were conducted at a central observing frequency of 1532\,MHz, with 768 channels across a bandiwidth of 384\,MHz. They were processed with the digital filterbank backend (DFB), where the data were incoherently dedispersed to the DM of the pulsar and folded at the best known period. The folded~\textit{archive} files have sub-integrations that are 10\,s long and 1024 phase bins. If a pulsar was not detected with the Lovell Telescope at 1532\,MHz, a dense observing campaign with the full LOFAR core was conducted. Four observations of 15 minutes each were conducted over a period of 10 days, with at least two of the observations over consecutive days. The data recording and processing followed the same setup used for the monthly timing campaign.

The folded \textit{archive} files obtained from the observations were analysed with \textsc{psrchive}. The pulse times-of-arrival (TOAs) of each pulsar were measured by cross-correlating the integrated pulse profiles of each observation with an analytical, noise free template. The templates were created by fitting between one and four von Mises functions\footnote{The von Mises function is a close approximation to a wrapped Gaussian function. See~\citet{ehp00} for a detailed description of the function.}, depending on the complexity of the pulse profiles~\citep{kwj+94}, with~\textsc{paas} of \textsc{psrchive}. An initial template was created by fitting the pulse profile of the observation with the highest signal-to-noise ratio (S/N) across the full LOFAR bandwidth to provide an interim timing solution. The LOFAR observations were then averaged into two frequency channels, with central frequencies of 128\,MHz and 167\,MHz in order to measure the $\mathrm{DM}$. Different templates were made for each observing frequency, taking into account possible profile evolution across the bandwidth~\citep{hsh+12}, fitting for the pulse profiles averaged across the observing span. The templates were then aligned at the half-maximum of the most prominent component, in order to take into account possible effects of scattering that might shift the peak of the profile at low frequencies to slightly later phase. For the isolated pulsars, only the average DM across the observing span is measured, without modelling the DM variation between individual observations. A phase-connected timing solution was then determined by modeling the TOAs with~\textsc{tempo2}~\citep{ehm06,hem06}. Any offsets between the TOAs obtained with LOFAR and the Lovell telescope were corrected by fitting a jump between the two sets of data. The Solar System ephemeris DE405 was used to model the motion of the bodies in the Solar System in order to convert the topocentric TOAs into barycentric TOAs.

As there are uncertainties in the initial spin periods of the pulsars from follow-up observations, there is an ambiguity in the number of pulsar rotations between successive monthly observations. In order to resolve the ambiguity and phase-connect the monthly TOAs, the TOAs from the dense campaign were used to model the rotation of the pulsars over short timescales. We were able to model the TOAs of all the pulsars with the TOAs from the dense observations with just a refinement of the spin period of the pulsars, producing coherent timing solutions that take into account all pulsar rotations within the short observing span. This allowed us to extrapolate the solutions to longer timescales without introducing ambiguities, which we then used to model the positions and spin period derivative of the pulsars.

\subsection{Timing of PSR J1658$+$3630} \label{sec:J1658timing}

The large apparent period derivative measured from the confirmation observation of PSR\,J1658$+$3630 strongly suggested that the pulsar is in a binary system. In order to model the binary orbit, more frequent observations of the pulsar are required, on top of the regular monthly observations. An initial dense observing campaign was first conducted to model the binary parameters of the system. The observing strategy is presented in Table~\ref{tab:1658obsstrategy} as Cam. 1.
Only the Stokes I data were recorded for these observations, with a central frequency of 149.8\,MHz and 6400 12.2\,kHz channels. The data were incoherently dedispersed to the DM of the pulsar and folded with the best known ephemeris at the time of the observations, producing \textit{archive} files with 10\,s sub-integrations and 256 phase bins.

\begin{table*}
    \centering
    \caption{Summary of the various observing strategies for timing PSR\,J1658$+$3630. The table shows the telescopes used, the observing campaigns, the span of the campaigns, the data acquisition methods, the observing strategies of each campaign and the observing lengths of individual observations within the campaigns.}
    \begin{tabular}{p{2cm}lllp{4.5cm}l}
    \hline
    Telescope & Obs. Campaign & Span & Data Acquisition & Obs. Strategy & Obs. length\\
    \hline
    LOFAR Core & Regular & 2017 Mar-2018 Nov & Complex Voltages & Monthly & 10 minutes\\
               & Cam. 1 Regular & 2017 Jan-Apr & Stokes I & Weekly & 5 minutes\\
               & Cam. 1 Dense 1 & 2017 Feb & Stokes I & 5 daily observations for 5 days & 5 minutes\\
               & Cam. 1 Dense 2 & 2017 Mar & Stokes I & 8 observations across 12 hours & 5 minutes\\
               & Cam. 2 Regular & 2017 Nov-2018 May & Complex Voltages & Weekly & 10 minutes\\
               & Cam. 2 Dense 1 & 2017 Nov & Complex Voltages & 6 observations across 12 hours around orbital phase 0.25 & 1 hour\\
               & Cam. 2 Dense 2 & 2017 Nov-2018 May & Complex Voltages & 5 observations at orbital phases of 0, 0.5 and 0.75 & 2 hour\\
    DE602, DE604, DE609 & Regular 1 & 2017 Jul-2018 May& Complex Voltages & Weekly & 2-3 hours\\
    DE602, DE603 & Regular 2 & 2018 Jun-Oct & Complex Voltages & Once every 2 weeks & 2-3 hours\\
    DE602, DE604, DE605 & Dense & 2017 Jul & Complex Voltages & 21 observations across 2 days & 2 hours\\
    \hline
    \end{tabular}
    \label{tab:1658obsstrategy}
\end{table*}

The binary orbit of PSR\,J1658+3630 was first determined by modelling the changes in the apparent spin period of the pulsar. The apparent spin periods at each epoch were measured by refolding the data at slightly different spin periods from the best known period and finding the spin period that produces the highest S/N using \textsc{pdmp} of \textsc{psrchive}. We then used the~\textsc{fitorbit} tool of~\textsc{psrtime}\footnote{\url{http://www.jb.man.ac.uk/pulsar/observing/progs/psrtime.html}} to model the apparent spin period of the pulsar at different epochs. We found that the variation of the apparent spin period is roughly sinusoidal, suggesting the pulsar is in an almost circular orbit and can be tentatively modelled assuming zero eccentricity. An initial timing solution was produced, modelling the orbital period, projected semi-major axis and the epoch of ascending node of the system and the spin period of the pulsar. This incoherent timing solution was then used to further refine the rotational, astrometric and binary properties of the pulsar using the timing procedure in Section~\ref{sec:timing} to obtain a coherent timing solution.

The incoherent timing solution of PSR\,J1658$+$3630 suggests a possibility of measurement of Shapiro delay of the pulsar signal due to its companion. 
Hence, a second dense campaign for PSR\,J1658$+$3630 was conducted to provide a large coverage of orbital phases, specifically at orbital phase 0.25, where the binary companion is in front of the pulsar. The observing strategy is described in Table~\ref{tab:1658obsstrategy} as Cam. 2.
These data were recorded, processed and analysed using the same method as the monthly timing observations, as described in Section~\ref{sec:timing}.

PSR\,J1658$+$3630 was also observed with five of the international LOFAR stations in Germany, namely stations in Unterweilenbach (telescope identifier DE602), Tautenburg (DE603), Bornim (DE604), J\"{u}lich (DE605) and Norderstedt (DE609), operated by the German LOng Wavelength consortium. The observing strategy and the individual stations involved are presented in Table~\ref{tab:1658obsstrategy}. The observations were conducted at a central frequency of 153.8\,MHz and with a bandwidth of 71.5\,MHz across 366 sub-bands. The data from the stations in Unterweilenbach, Tautenburg and J\"{u}lich were recorded on machines at the Max-Planck-Institut f\"{u}r Radioastronomie in Bonn, while the data from stations in Bornim and Norderstedt were record at machines in J\"{u}lich Supercomputing Centre. They were recorded using the LOFAR und MPIfR Pulsare (LuMP4) software\footnote{\url{https://github.com/AHorneffer/lump-lofar-und-mpifr-pulsare}} as channelized complex voltages and then coherently dedispersed to the DM of the pulsar and folded using the best ephemeris of the pulsar available in 2017 July, producing \textit{archive} files with ten-second sub-integrations and 1024 phase bins. A summary of the different observing strategies on PSR\,J1658$+$3630 is shown in Table~\ref{tab:1658obsstrategy}.

\subsection{Pulse Profile and Spectral Properties Analysis}

The timing observations conducted with both LOFAR and the Lovell Telescope were used to study the pulse profile evolution and the spectral properties across different frequencies. Additionally, the pulsars were observed at a central observing frequency of 334\,MHz with the Lovell Telescope to provide a larger frequency coverage in these studies. Each pulsar was observed once for 30 minutes. These data have a bandwidth of 64\,MHz across 512 frequency channels and were processed with the DFB. The folded~\textit{archive} files have sub-integrations of 10\,s and 512 phase bins.

The pulse profiles of the pulsars at 149 and 1532\,MHz were obtained by adding the integrated pulse profiles obtained from each observation together with \textsc{psradd} of \textsc{psrchive}, after aligning the profiles from different observations using the timing solutions presented in Table~\ref{tab:ephemeris}. The pulse profiles at 334\,MHz were obtained from the single observations made. We measured the pulse widths and the duty cycles of the pulsars at 10 and 50 per cent of the maximum, W$_{10}$, $\delta_{10}$, W$_{50}$, $\delta_{10}$, respectively, using the following method. First, noise free templates of the pulse profiles were generated by fitting a number of von Mises functions, each with a different height, width and position using \textsc{paas}. The templates formed are the summation of individual von Mises functions that best represents the overall shapes of the profiles. The observed off-pulse region was then used to generate a noise distribution. One thousand simulated pulses were generated by creating new noisy profiles combining the noise free template with the simulated noise. These were then fitted with the noise free templates, allowing height, width and position of each individual function to vary. The widths of the remodelled templates at 10 and 50 per cent of the maximum from the 1000 trials were then used to determine the mean width and error. The widths are defined as the outermost components of the templates that are at 10 per cent and above, and 50 per cent and above the maximum.

The LOFAR observations were flux-calibrated using the method described in~\citet{kvh+16} in order to measure the flux densities of the pulsars. The calibrated data of individual observations were then averaged into two frequency bands to measure the flux densities of the pulsars at 128 and 167\,MHz respectively. The average flux density of each pulsar was then computed from these observations. We found that the offset between the initial pointing position of the observations of most of the pulsars and the measured position obtained through pulsar timing is of the order of one arcmin. The pointing offset was corrected for several of the pulsars studied here in later observations. We attempted to correct the measured flux densities due to loss in sensitivity as a result from the offsets from the early observations. However, we found that pulsars with initial pointing positions that are offset from the timing position by less than 1 arcmin did not show any improvement in S/N after updating the pointing position, while those with offset of more than 1 arcmin showed S/N improvement that is less than expected. We suspect that this could be due to two issues. First, the ionosphere is known to induce a jitter in the position of the pulsar in the sky up to 1 arcmin around the actual position of the source. We also suspect that the shape of the tied-array beam formed cannot be modelled as a 2D-Gaussian due to the complexity of the beam shape and the beam not being fully coherent. Hence, we decided that, for pulsars with an initial pointing offset less than one arcmin, flux density measurements for all observations were used without applying any correction, while for those with larger offsets, only the observations after the correction of pointing position were used to measure the flux density of the pulsars. The error in the measured flux densities of each frequency band of a single observation is conservatively estimated to be 50 per cent~\citep{bkk+16,kvh+16}.

The flux densities of pulsars detected at 334 and 1532\,MHz were estimated using the radiometer equation~\citep{lk05}. The receiver temperature T$_{\mathrm{rec}}$ of the 334\,MHz receiver is 50\,K and the sky temperature T$_{\mathrm{sky}}$ in the direction to the relevant pulsar was estimated by extrapolating the T$_{\mathrm{sky}}$ at 408\,MHz~\citep{hssw82} with a spectral index of $-2.55$~\citep{lmop87,rr88}. The Gain, G, of the receiver is 1 K Jy$^{-1}$ and the bandwidth was 64\,MHz. As for the 1532\,MHz observations, T$_{\mathrm{rec}} = 25\,\mathrm{K}$, G$=1\,\mathrm{K}\,\mathrm{Jy}^{-1}$ and the bandwidth was 384\,MHz. We estimated the average RFI fraction of each observation at 1532\,MHz to be 20 per cent of the bandwidth and 5 per cent of the observing length, while at 334\,MHz, the fraction is 20 per cent of the bandwidth and 20 per cent of the observing length. Upper limits for the non-detections were estimated using the radiometer equation with a threshold S/N of 10 and estimated pulse width based on the measured W$_{50}$ at 149\,MHz. The uncertainty on the flux density of a single observation is estimated to be 20 per cent.

\section{Timing Properties of Isolated Pulsars} \label{sec:timingprop}

The timing solutions of the 20 isolated pulsars are shown in Table~\ref{tab:ephemeris}, and the derived properties based on the rotational parameters are shown in Table~\ref{tab:derived}. The timing residuals of the 20 pulsars using the best solutions are shown in Fig.~\ref{fig:residuals1}. The locations of the pulsars in the $P$-$\dot{P}$-diagram are shown in Fig.~\ref{fig:LOTAAStimingppdot}.

\begin{table*}
\centering
\caption{The timing solutions of 20 pulsars, showing the positions, reference epoch, spin period, spin-period derivative, DM, number of TOAs used to model the pulsars and the timing residuals after modelling the pulsars.}
\label{tab:ephemeris}
\begin{tabular}{lllllllllll}
\hline
PSR & RA & DEC & Epoch & P & $\dot{\text{P}}$ & DM & N$_\mathrm{TOA}$ & TRES & $\chi^2_{\textrm{red}}$\\
(J2000) & (J2000) & (J2000) & (MJD) & (s) & (10$^{-15}$) & (pc\,cm$^{-3}$) &  & ($\upmu$s) & \\
\hline
J0100$+$8023 &	01:00:16.09(1) &	$+$80:23:41.75(3) &	56676 &	1.4936009186(2) &	0.3543(6) &	56.0062(8) & 40 & 242 & 0.73\\
J0107$+$1322 &	01:07:39.95(2) &	$+$13:22:31.7(8) &	57972 &	1.1973833938(1) &	0.687(3) &	21.671(1) & 34 & 331 & 1.3\\
J0210$+$5845 &	02:10:55.4(1) &	$+$58:45:04(1) &	58011 &	1.766208099(2) &134.02(2) &	76.772(6) & 32 & 1440 & 3.6\\
J0421$+$3255 &	04:21:33.2(2) &	$+$32:55:50(10) &	57310 &	0.900105016(1) &0.06(2) &	77.02(6) & 50 & 20744 & 24\\
J0454$+$4529 &	04:54:59.310(8) &	$+$45:29:46.7(1) &	56909 &	1.3891369360(4) &	4.888(2) &	20.834(2) & 36 & 408 & 0.92\\
J1017$+$3011 &	10:17:36.29(3) &	$+$30:11:46.1(4) &	57986 &	0.4527850730(2) &	0.57(3) &	27.150(2) & 30 & 561 & 0.89\\
J1624$+$5850 &	16:24:00.964(9) &	$+$58:50:15.77(6) &	57718 &	0.65180081878(2) &	0.331(1) &	26.403(2) & 50 & 617 & 1.3\\
J1638$+$4005 &	16:38:16.243(6) &	$+$40:05:56.37(6) &	57491 &	0.76772039193(9) &	0.183(2) &	33.417(1) & 40 & 435 & 0.98\\
J1643$+$1338 &	16:43:54.140(5) &	$+$13:38:43.9(1) &	57820 &	1.09904716266(7) &	0.778(2) &	35.821(1) & 45 & 379 & 1.6\\
J1656$+$6203 &	16:56:10.29(1) &	$+$62:03:50.41(9) &	57652 &	0.77615531125(4) &	0.844(2) &	35.262(3) & 52 & 828 & 0.80\\
J1657$+$3304 &	16:57:50.682(6) &	$+$33:04:33.65(5) &	57927 &	1.5702755247(1) &	1.734(2) &	23.9746(6) & 40 & 191 & 1.2\\
J1713$+$7810 &	17:13:27.07(3) &	$+$78:10:33.99(8) &	57773 &	0.43252593524(3) &	0.114(5) &	36.977(3) & 42 & 810 & 1.1\\
J1741$+$3855 &	17:41:12.341(8) &	$+$38:55:09.90(6) &	57927 &	0.82886088996(6) &	0.206(4) &	47.224(2) & 40 & 495 & 1.1\\
J1745$+$1252 &	17:45:44.19(2) &	$+$12:52:38.3(2) &	57883 &	1.0598487584(1) &	0.563(4) &	66.141(5) & 46 & 1434 & 1.1\\
J1749$+$5952 &	17:49:33.228(2) &	$+$59:52:36.13(1) &	57751 &	0.436040950719(2) &	0.1519(4) &	45.0694(4) & 58 & 134 & 1.1\\
J1810$+$0705 &	18:10:47.038(7) &	$+$07:05:36.3(2) &	57770 &	0.30768283388(1) &	0.242(4) &	79.425(5) & 59 & 1045 & 2.6\\
J1916$+$3224 &	19:16:03.468(4) &	$+$32:24:39.70(6) &	57638 &	1.13744972551(3) &	3.4696(6) &	84.105(2) & 56 & 565 & 1.9\\
J1957$-$0002 &19:57:42.620(7) &	$-$00:02:06.8(2) &	57674 &	0.96509596606(4) &	0.805(1) &	38.443(4) & 39 & 621 & 0.93\\
J2036$+$6646 &	20:36:52.32(5) &	$+$66:46:20.7(3) &	57974 &	0.5019271782(2) &	0.94(4) &	50.763(2) & 30 & 607 & 1.6\\
J2122$+$2426 &	21:22:39.02(1) &	$+$24:26:44.9(2) &	57922 &	0.54142115903(4) &	0.127(6) &	8.500(5) & 47 & 1029 & 1.4\\
\hline
\end{tabular}
\end{table*}

\begin{table}
\centering
\caption{The derived properties of the 20 pulsars based on the rotational parameters obtained, showing the characteristic age, $\log{\tau}_\text{c}$, surface magnetic field strength, $\log\text{B}$, and the rotational energy loss, $\log\dot{\text{E}}$. The derived quantities assume that the pulsars have a dipolar magnetic field structure (See~\citealt{lk05} for the descriptions of the derived quantities.)}
\label{tab:derived}
\begin{tabular}{llll}
\hline
PSR & $\log{\tau}_\text{c}$ & $\log\text{B}$ & $\log\dot{\text{E}}$\\
 & (yr) & (G) & (erg\,s$^{-1}$)\\
\hline
J0100$+$8023 &	7.8 &	11.9 &	30.6 \\
J0107$+$1322 &	7.4 &	12.0 &	31.2 \\
J0210$+$5845 &	5.3 &	13.2 &	33.0 \\
J0421$+$3255 &	8.3 &	11.4 &	30.5 \\
J0454$+$4529 &	6.7 &	12.4 &	31.9 \\
J1017$+$3011 &	7.1 &	11.7 &	32.4 \\
J1624$+$5850 &	7.5 &	11.7 &	31.7 \\
J1638$+$4005 &	7.8 &	11.6 &	31.2 \\
J1643$+$1338 &	7.4 &	12.0 &	31.4 \\
J1656$+$6203 &	7.2 &	11.9 &	31.9 \\
J1657$+$3304 &	7.2 &	12.2 &	31.2 \\
J1713$+$7810 &	7.8 &	11.4 &	31.7 \\
J1741$+$3855 &	7.8 &	11.6 &	31.2 \\
J1745$+$1252 &	7.5 &	11.9 &	31.3 \\
J1749$+$5952 &	7.7 &	11.4 &	31.9 \\
J1810$+$0705 &	7.3 &	11.4 &	32.5 \\
J1916$+$3224 &	6.7 &	12.3 &	32.0 \\
J1957$-$0002 &	7.3 &	12.0 &	31.5 \\
J2036$+$6646 &	6.9 &	11.8 &	32.5 \\
J2122$+$2426 &	7.8 &	11.4 &	31.5 \\
\hline
\end{tabular}
\end{table}

\begin{figure*} 
\includegraphics[width=\linewidth]{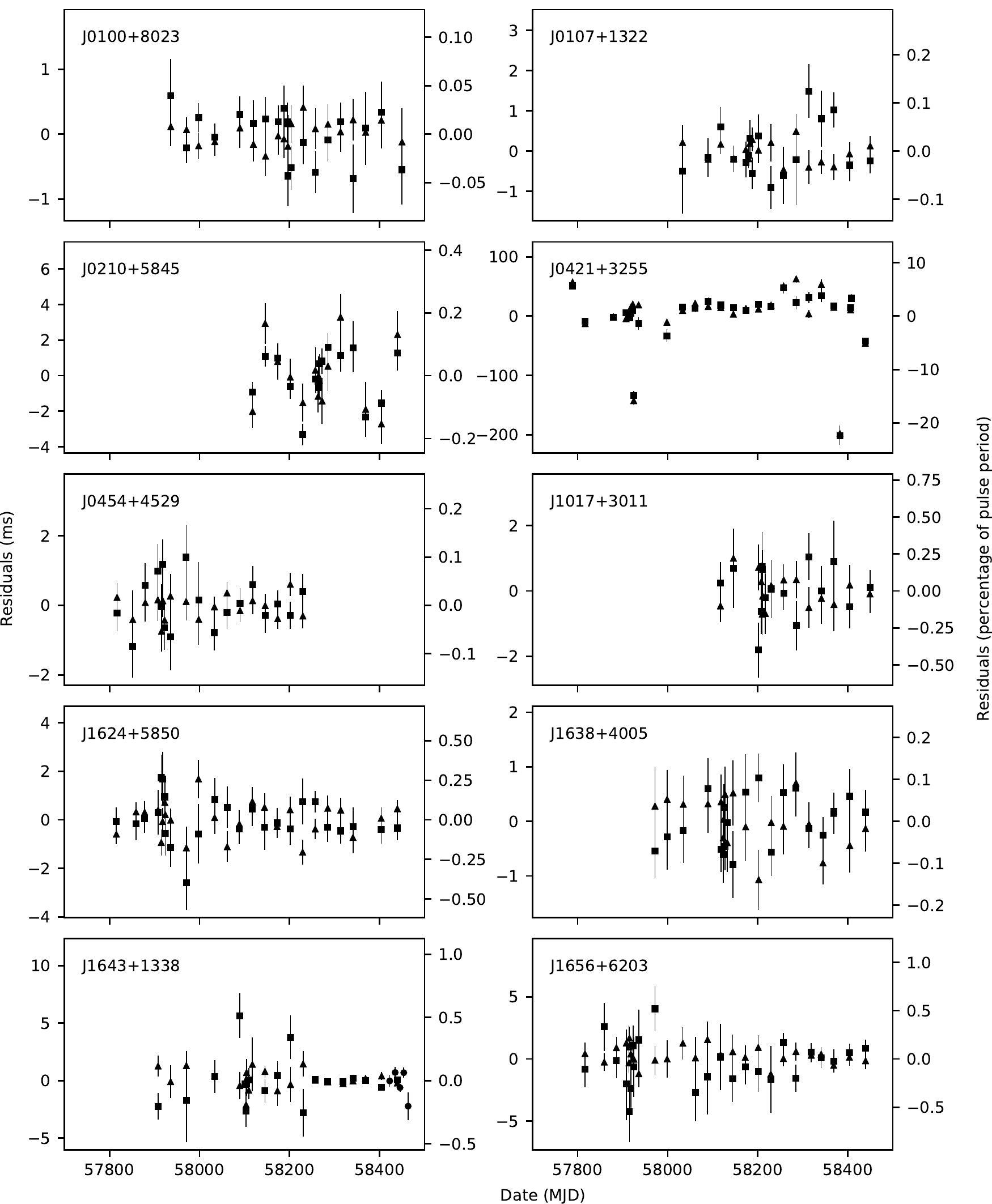}
\caption{The residuals from the timing model of the TOAs from the 20 pulsars presented in Table~\ref{tab:ephemeris}. The different symbols represents the different observing frequencies, with triangles for the lower part of the LOFAR band (128\,MHz), squares for the upper part of the LOFAR band (167\,MHz) and dots for the Lovell telescope at 1532\,MHz.}
\label{fig:residuals1}
\end{figure*}

\setcounter{figure}{0}
\begin{figure*} 
\includegraphics[width=\linewidth]{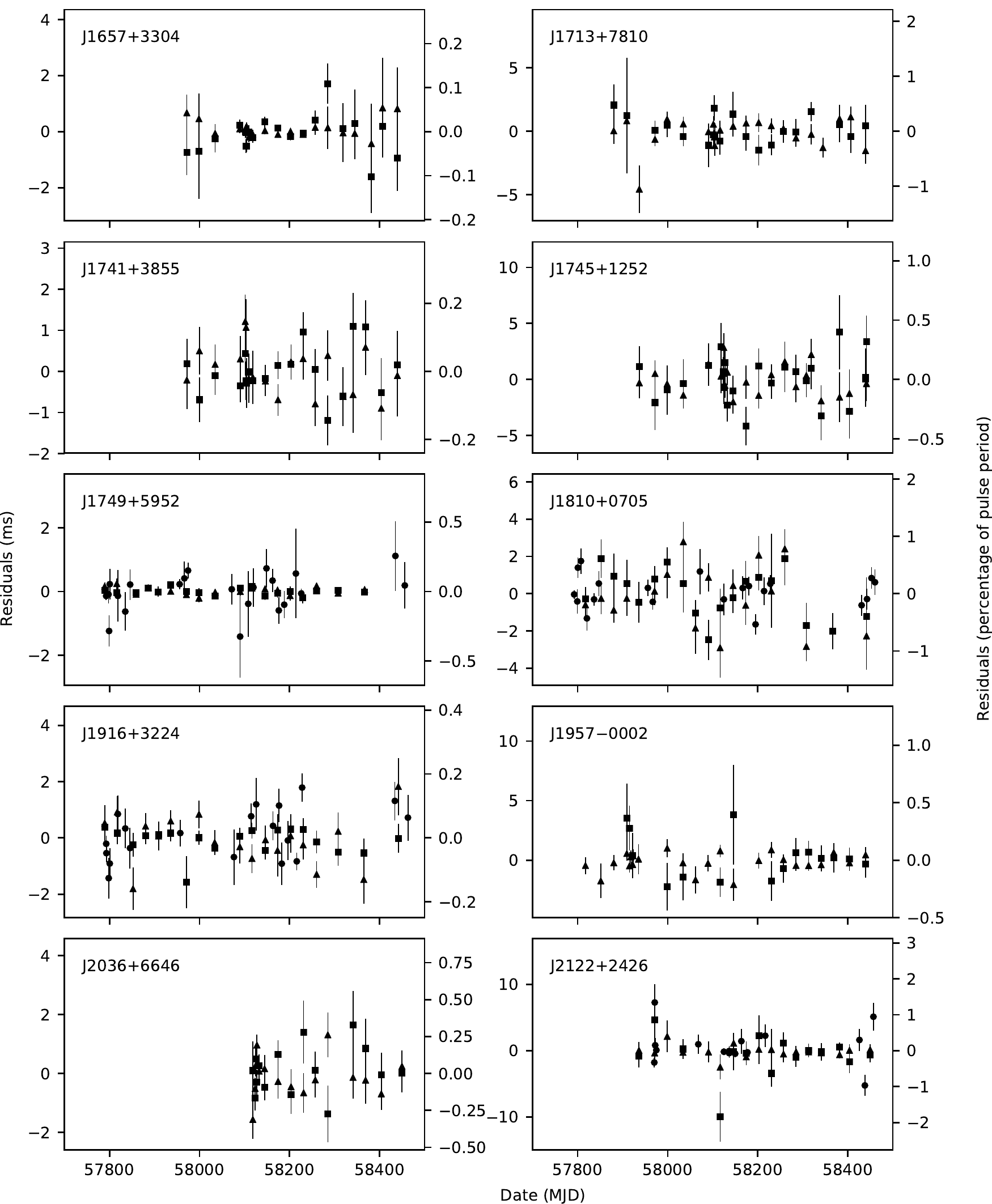}
\caption{Continued.}
\label{fig:residuals2}
\end{figure*}

\begin{figure}
\centering
\includegraphics[width=\linewidth]{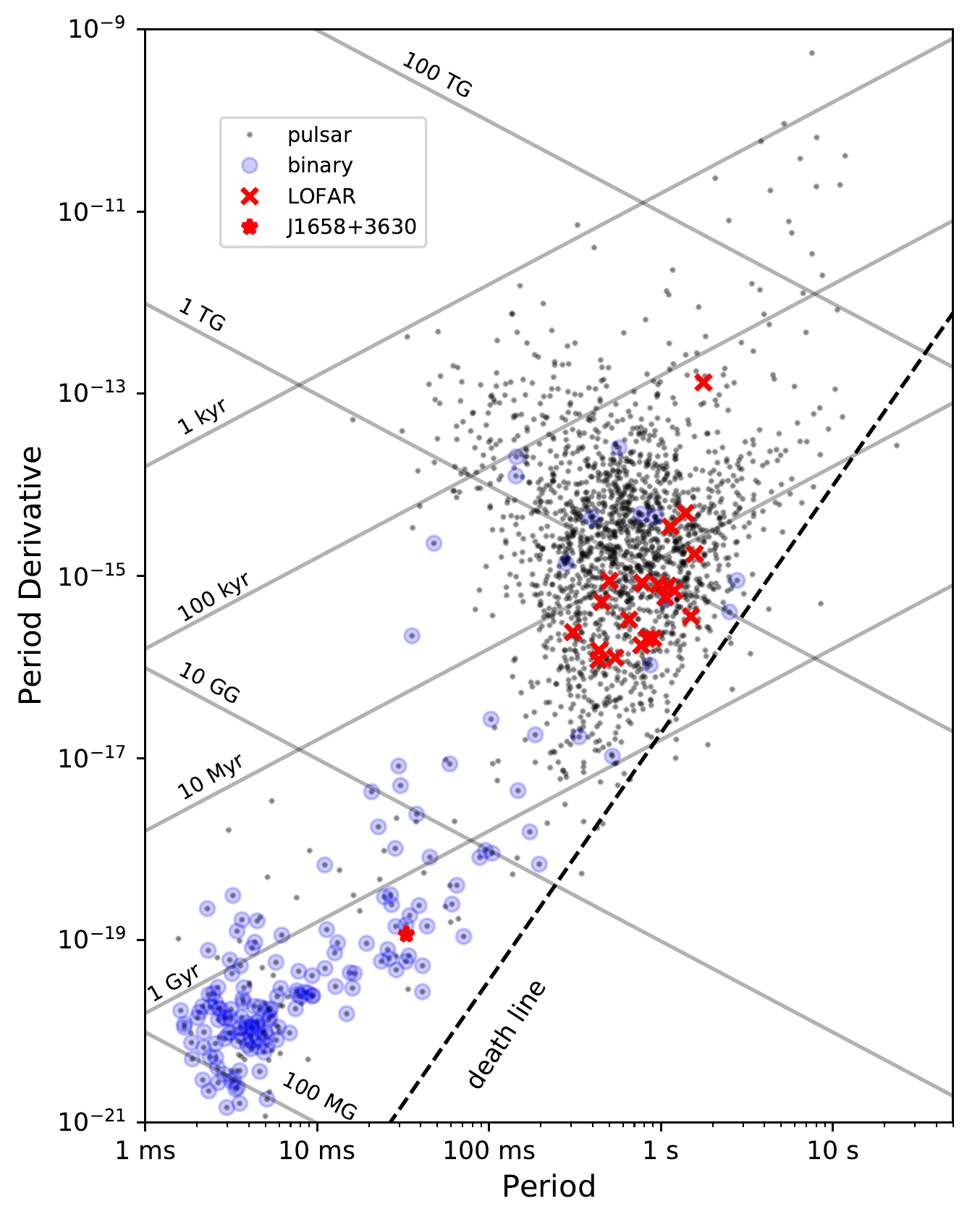}
\caption{The $P$-$\dot{P}$-diagram of pulsars, overlaid with the 21 pulsars being studied here. The 20 isolated pulsars are indicated with a red cross and PSR\,J1658$+$3630 indicated with a red star. The death line plotted here is the one modelled by Equation 9 of~\citet{cr93}. Lines of constant characteristic ages and inferred surface magnetic field strengths are indicated in the plot.}
\label{fig:LOTAAStimingppdot}
\end{figure}

The properties of most of the pulsars presented here are well modelled with just a period, a period derivative, and the position. However, PSRs\,J0210$+$5845 and J0421$+$3255 showed large timing residuals, as shown by their relatively large $\chi^2_{\textrm{red}}$ values. The timing residuals of PSR\,J0210$+$5845 suggest that the pulsar has large timing noise related to its small characteristic age of about 200 kyr, which requires extra frequency derivatives to model the TOAs that could be determined through longer observing span. PSR\,J0421$+$3255 has a wide pulse profile which increases the uncertainties of individual TOAs. Furthermore, two of the observations of PSR\,J0421$+$3255 have TOAs that are offset compared to the rest of the observations by roughly 20\% of a pulse phase. These observations are not used to model the properties of PSR\,J0421$+$3255, as we later found that the pulses from these observations coincide with a change in the integrated pulse profile (See Section~\ref{sec:20pulseprofile} for further discussion).

PSR\,J1643$+$1338 was detected at low S/N in its initial timing observations. After 10 months of observations, the timing position of the pulsar was found to be different from the observed position by 13 arcmin. The corrected position was used from 2018 May onwards, with the pulsar detected at a S/N about 6 times higher than before. Although the position offset of the pulsar is roughly 4 times the 3.5-arcmin full-width half-maximum (FWHM) of the LOFAR core beam at 149\,MHz, the relatively high brightness of the pulsar allowed us to detect it in a side lobe of the main beam.

The TOA uncertainty of individual observations of PSR\,J1657$+$3304 varied by about a factor of 20 across the observation span. This corresponds to a large variation in the observed flux density in the pulsar, with S/N changes of a factor of 10. Further analysis and discussion of PSR\,J1657$+$3304 are presented in Section~\ref{sec:J1657}.

The 20 LOTAAS discoveries discussed in this section are indicated by small red crosses in Fig.~\ref{fig:LOTAAStimingppdot}. Most of the pulsars studied here are located in the lower right of the $P$-$\dot{P}$-diagram, where the characteristic ages are $\tau_{\text{c}} \gtrsim 10$ Myr. This suggests that the population that is probed here is older than the general pulsar population. Qualitatively, this fits with the expectations that LOFAR surveys are in general limited by scatter broadening~\citep{ls10}. In combination with dispersive effects, LOTAAS is more sensitive to longer-period sources than existing, higher-frequency surveys. Hence, discoveries from LOTAAS will generally be longer period and, on average, older. However, this sample only represents a fraction of all the pulsars discovered by LOTAAS. A broader study on the properties of all LOTAAS discoveries will be conducted after the conclusion of the survey.

\subsection*{PSR J0210$+$5845 -- small $\tau_{\text{c}}$, high B pulsar}

A notable exception to this sample of pulsars is PSR\,J0210$+$5845. It is located in a more sparsely populated region of the diagram, with a characteristic age of $\tau_{\text{c}} \approx 200$\,kyr and surface dipole magnetic field of $\text{B} \approx 1.1\times 10^{13}$\,G. The pulsar also shows large timing residuals, which could be timing noise related to its small characteristic age~\citep{hlk10}. Additionally, the large timing residual suggest that high order frequency derivatives are potentially measurable in the future with a larger observing span of a few years.

\section{Pulse Profiles of Isolated Pulsars} \label{sec:20pulseprofile}

\begin{figure*}
\includegraphics[width=\linewidth]{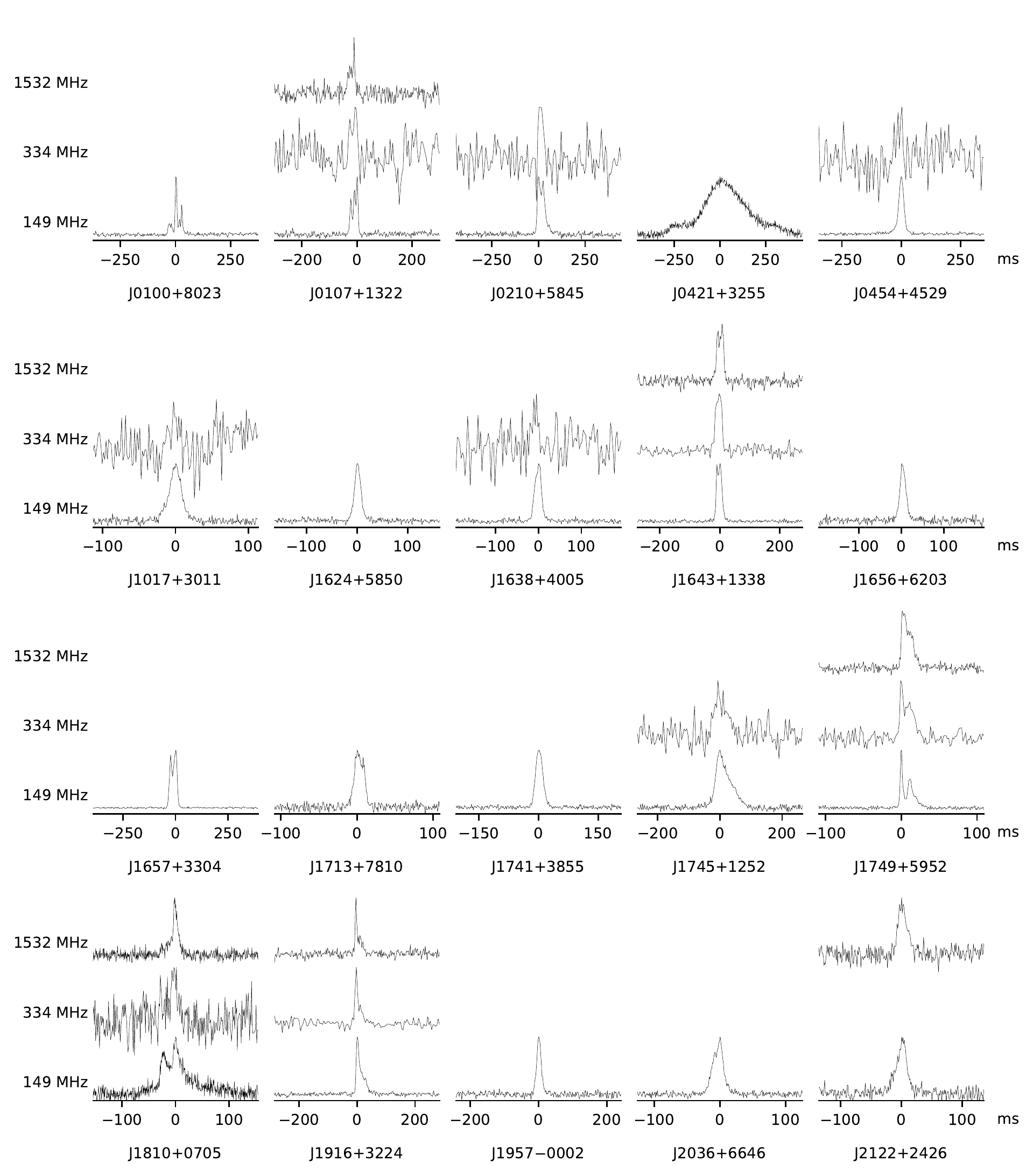}
\caption{The pulse profiles at 149, 334 and 1532\,MHz for the 20 pulsars shown in Table~\ref{tab:ephemeris}. For each profile, 50 per cent of the pulse phase is shown, except for PSRs\,J0421+3255 and J1810$+$0705, where the full pulse phase is shown. The profiles at 149 and 1532\,MHz have 512 profile bins across the pulse phase, while those at 334\,MHz have 256 profile bins across the pulse phase. The scales provide a reference to the width of the profiles in milliseconds.}
\label{fig:profiles}
\end{figure*}

The pulse profiles of the 20 isolated pulsars studied here at 149, 334 and 1532\,MHz are presented in Fig.~\ref{fig:profiles}. The pulse widths of the pulsars at these frequencies are presented in Table~\ref{tab:widths}.

Amongst the 20 pulsars studied here, nine are only detected with LOFAR at 149\,MHz. Six of these, PSRs\,J1624$+$5850, J1656$+$6203, J1713$+$7810, J1741$+$3855, J1957$-$0002 and J2036$+$6646, show single peaks in their integrated pulse profiles, four of which are well modelled with just a single von Mises component. The profiles of PSRs\,J1713$+$7810 and J2036$+$6646 are more complex and require two components to model. The profile of PSR\,J1657$+$3304 shows two distinct peaks, both of which are well modelled with just a single component. The profile of PSR\,J0100$+$8023 is the most complex of all the pulsars studied here. It consists of four separate peaks. Two separate values of W$_{50}$ were measured for PSR\,J0100$+$8023: 8.01\,ms, which corresponds to the width of the main peak, and 30.4\,ms, which is the combined width of the main peak and the fourth peak at the trailing edge of the profile. This is due to the relative intensity of this outer peak is about half of the main peak, which results in some of the measurements included this peak and others not. This is also seen in three other measurements of pulse widths. 

\begin{figure}
  \centering
  \includegraphics[width=\linewidth]{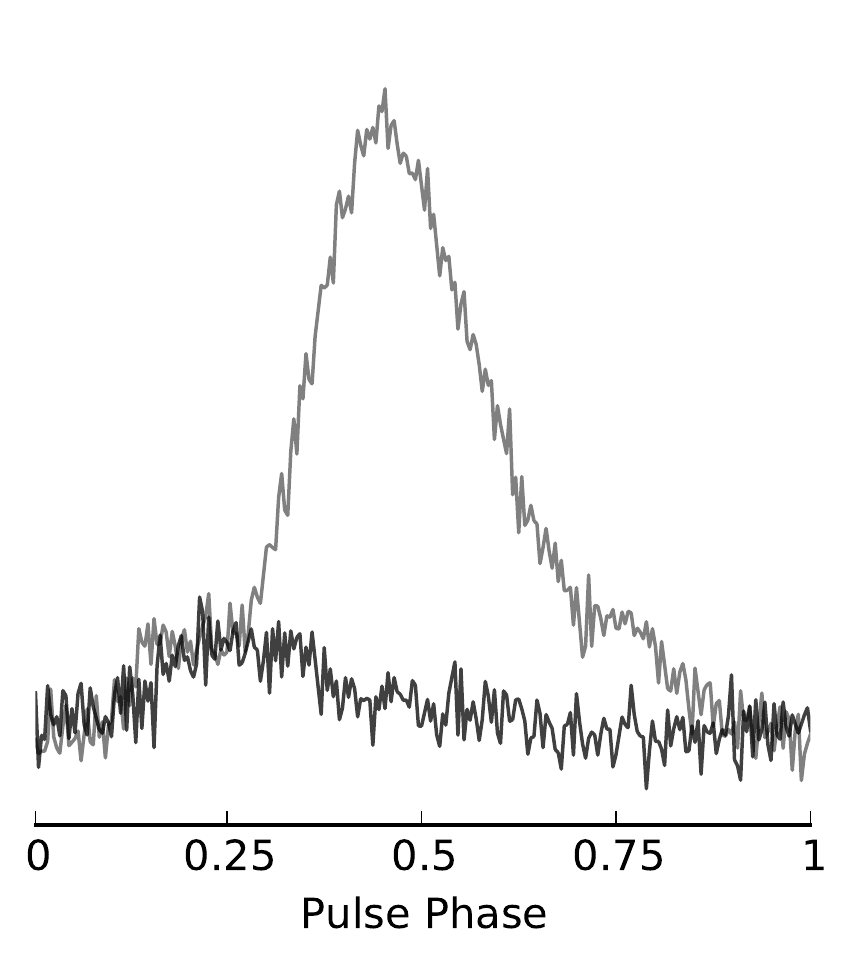}
  \caption{The integrated pulse profile of PSR\,J0421$+$3255 (grey) created by summing all available observations, except the two that show a timing offset (see Fig.~\ref{fig:residuals1}). Overlaid in black is the integrated pulse profile derived from just the two observations showing a timing offset. The difference in profile shapes  suggests that there is mode changing, where the pulsar emission sometimes occurs preferentially at the leading edge of the regular emission. The integrated pulse profiles are normalized via the off-pulse standard deviation between phases 0.9 and 1.}
  \label{fig:J0421modes}
\end{figure}

The profile of PSR\,J0421$+$3255 showed a main peak with intensity that is much larger than two smaller peaks at either side of it. The location of the leading component of PSR\,J0421$+$3255 coincides with the position of the pulse of the two observations that have a timing offset, shown in Fig.~\ref{fig:J0421modes}. As the pulse profile is formed without adding these offset observations, this suggests that the pulsar might undergo mode changing where there is only emission from the leading peak, which has larger intensity compared to the regular mode in these observations. Alternatively, the profile change could be similar to the flaring phenomenon seen in PSR\,B0919$+$06~\citep[e.g.][]{rrw06,psw+15}, where the emission of the pulsar appears to shift earlier in pulse phase compared to the regular emission. However, the flares seen in PSR\,B0919$+$06 last on the order of seconds while in PSR\,J0421$+$3255, they last longer than a single 15 minutes observation.

Five of the pulsars are detected with both LOFAR at 149\,MHz and the Lovell telescope at 334\,MHz, but not at 1532\,MHz. PSRs\,J0454$+$4529, J1017$+$3011, J1638$+$4005 and J1745$+$1252 show complex single peaked profiles that are modelled with multiple components at 149\,MHz. The detections at 334\,MHz for PSRs\,J0454$+$4529, J1017$+$3011 and J1638$+$4005 are too weak for the profiles to be well modelled. However, the general shapes of the profiles are similar to those at 149\,MHz.

\begin{landscape}
\begin{table}
\centering
\caption{The pulse widths and duty cycles of the pulsars shown in Fig~\ref{fig:profiles}. Four measurements of $\text{W}_{50}$ and $\delta_{50}$ showed two distinct sets of values corresponding to whether the measurements includes outer components with height that are at about 50 per cent of maximum. See Section~\ref{sec:20pulseprofile} for more details.}
\label{tab:widths}
\begin{tabular}{lcccccccccccccccc}
\hline
PSR && \multicolumn{3}{c}{$\text{W}_{10}$ (ms)} && \multicolumn{3}{c}{$\delta_{10}$ (\%)} && \multicolumn{3}{c}{$\text{W}_{50}$ (ms)} && \multicolumn{3}{c}{$\delta_{50}$ (\%)} \\
  & & 149 & 334 & 1532  & & 149 & 334 & 1532  & & 149 & 334 & 1532  & & 149 & 334 & 1532 \\
 & & MHz & MHz & MHz  & & MHz & MHz & MHz  & & MHz & MHz & MHz  & & MHz & MHz & MHz \\
\hline
J0100$+$8023 & & 67(3)& --& -- & & 4.5(2)& --& -- & & 8.01(1)$^a$& --& -- & & 0.5361(9)$^a$& --& -- \\
J0107$+$1322 & & 35.7(2)& 44(2)& 39(4) & & 2.98(2)& 3.7(1)& 3.2(3) & & 28.0(3)& 32(2)& 3.7(2)$^b$ & & 2.34(3)& 2.7(2)& 0.31(2)$^b$ \\
J0210$+$5845 & & 76(3)& 49(6)& -- & & 4.3(2)& 2.8(2)& -- & & 40.9(6)& 27(3)& -- & & 2.31(3)& 1.5(2)& -- \\
J0421$+$3256 & & 634(9)& --& -- & & 70(1)& --& -- & & 231(2)& --& -- & & 25.6(2)& --& -- \\
J0454$+$4529 & & 48.3(8)& 67(3)& -- & & 3.48(5)& 4.8(2)& -- & & 23.54(9)& 37(3)& -- & & 1.695(7)& 2.6(2)& -- \\
J1017$+$3011 & & 38.8(6)& 30.3(9)& -- & & 8.6(1)& 6.7(2)& -- & & 18.9(2)& 16.6(9)& -- & & 4.17(5)& 3.7(2)& -- \\
J1624$+$5850 & & 27.0(3)& --& -- & & 4.14(5)& --& -- & & 14.8(3)& --& -- & & 2.27(5)& --& -- \\
J1638$+$4005 & & 29.7(2)& 33(2)& -- & & 3.87(2)& 4.3(2)& -- & & 18.5(2)& 18(2)& -- & & 2.41(3)& 2.4(2)& -- \\
J1643$+$1338 & & 31.0(1)& 33.6(5)& 35.9(3) & & 2.823(9)& 3.06(5)& 3.27(3) & & 20.8(1)& 23.4(6)& 25.8(4) & & 1.89(1)& 2.13(6)& 2.34(4) \\
J1656$+$6203 & & 30.4(4)& --& -- & & 3.91(5)& --& -- & & 16.7(4)& --& -- & & 2.15(5)& --& -- \\
J1657$+$3304 & & 49.8(1)& --& -- & & 3.171(7)& --& -- & & 38.1(1)& --& -- & & 2.426(8)& --& -- \\
J1713$+$7810 & & 24(1)& --& -- & & 5.6(3)& --& -- & & 16(1)& --& -- & & 3.7(2)& --& -- \\
J1741$+$3855 & & 38.7(4)& --& -- & & 4.67(5)& --& -- & & 21.2(4)& --& -- & & 2.55(5)& --& -- \\
J1745$+$1252 & & 105.4(6)& 85(4)& -- & & 9.95(6)& 8.0(4)& -- & & 47(2)& 48(6)& -- & & 4.4(1)& 4.6(6)& -- \\
J1749$+$5952 & & 26.1(3)& 26.6(5)& 25.1(2) & & 5.99(6)& 6.1(1)& 5.75(5) & & 13.3(2)$^c$& 14.5(11)$^d$& 14.7(4) & & 3.04(4)$^c$& 3.3(3)$^d$& 3.38(9) \\
J1810$+$0705 & & 127(12)& 46(6)& 36(1) & & 41(4)& 15(2)& 11.6(4) & & 42.3(14)& 20(5)& 9.4(2) & & 13.8(4)& 6.4(16)& 3.05(5) \\
J1916$+$3224 & & 46.6(4)& 37(2)& 40(2) & & 4.10(3)& 3.2(2)& 3.5(1) & & 11.2(1)& 7.9(3)& 5.2(1) & & 0.99(1)& 0.69(2)& 0.461(8) \\
J1957$-$0002 & & 28.4(5)& --& -- & & 2.95(5)& --& -- & & 15.6(5)& --& -- & & 1.62(5)& --& -- \\
J2036$+$6646 & & 31.8(2)& --& -- & & 6.33(4)& --& -- & & 19.5(3)& --& -- & & 3.88(6)& --& -- \\
J2122$+$2426 & & 41(1)& --& 35(1) & & 7.7(2)& --& 6.5(2) & & 19.5(7)& --& 15.5(8) & & 3.6(1)& --& 2.9(1) \\
\hline
\multicolumn{17}{l}{$^a$Also measured as 30.4(3)\,ms and 2.04(2)}\\
\multicolumn{17}{l}{$^b$Also measured as 18(2)\,ms and 1.5(2)}\\
\multicolumn{17}{l}{$^c$Also measured as 3.04(1)\,ms and 0.697(3)}\\
\multicolumn{17}{l}{$^d$Also measured as 4.3(1)\,ms and 0.98(2)}\\
\end{tabular}
\end{table}
\end{landscape}

The profiles of PSRs\,J0210$+$5845, J1745$+$1252 and J1810$+$0705 show signs of a possible scattering tail. The scattering tail is modelled as an exponential function with a scale $\tau_s$ known as the scattering timescale. $\tau_s$ has a power law dependence on observing frequency: $\tau_s \propto {\nu}^{\alpha_s}$, where ${\alpha_s}$ is known as the scattering index. We measured the scattering index, $\alpha_s$ and timescale, $\tau_s$ of PSRs\,J0210$+$5845 and J1745$+$1252 across the LOFAR band using the method described in~\citet{gk16}, assuming an isotropic scattering screen, where the scattered photons is described as a circularly symmetric Gaussian distribution around the source. The scattering properties of PSR\,J1810$+$0705 are not modelled as the method used currently does not support profiles that cannot be described by a single Gaussian component. The LOFAR band is split into 8 sub-bands and the scattering time in each band is measured, with $\alpha_s$ fitted assuming a power law relationship between scattering time and frequency. The measured $\alpha_s$ and $\tau_s$ at 150\,MHz of PSR\,J0210$+$5845 are $-2.4 \pm 0.5$ and $19 \pm 2$ ms respectively, while the measured $\alpha_s$ and $\tau_s$ at 150\,MHz of PSR\,J1745$+$1252 are $-1.0 \pm 0.1$ and $33 \pm 1$ ms respectively. The measured scattering indices of the two pulsars are much lower than the theoretically predicted relationship of $\alpha_s = -4$~\citep{cro70,lan71} or $-4.4$~\citep{lj76,ric77}. However, this was seen on other pulsars observed with LOFAR~\citep{lmg+04,gkk+17}, which can be explained by the finite size of the scattering screen. Alternatively, the low scattering indices measured, especially for PSR\,J1745$+$1252, could be due to intrinsic profile evolution that mimicked a scattering tail.

PSR\,J2122$+$2426 is detected with LOFAR at 149\,MHz and the Lovell telescope at 1532\,MHz. The pulsar shows a single-peaked profile that is modelled with 2 components at both frequencies. However, the positions of the second component are different at the two observing frequencies, where it appears before the main component at 149\,MHz but after the main component at 1532\,MHz. This evolution is similar to what is observed with PSR\,B0809$+$74 by~\citet{hsh+12}. However, the observation of the pulsar at 334\,MHz is strongly affected by RFI and so we cannot be sure if this is the same behaviour.

Five pulsars -- PSRs\,J0107$+$1322, J1643$+$1338, J1749$+$5952, J1810$+$0705 and J1916$+$3224 -- are detected with LOFAR and the Lovell telescope at both 334 and 1532\,MHz. These detections allowed us to study the evolution of the pulse profiles across multiple different observing frequencies. Here we describe the frequency evolution of these pulsars one by one.

\subsection*{PSR\,J0107$+$1322 -- triple peaked profile}

PSR\,J0107$+$1322 shows three distinct peaks in its profile at 149\,MHz, which are well modelled with three components that increase in intensity from the leading to the trailing peaks. However, the profiles at 334 and 1532\,MHz only show two distinct peaks; a weak leading peak and a strong trailing peak. The profiles are better modelled with two components rather than with three components. It is possible that the profiles at higher frequencies has 3 separate components which are not visible due to low S/N of the observations. While the increase in W$_{10}$ from 149 to 334\,MHz suggest a profile evolution contrary to the RFM model, higher S/N observations of PSR\,J0107$+$1322 at higher frequencies are required to confirm the relationship, as the low S/N observation at 334\,MHz could also introduce an extra uncertainty to the determination of the off-pulse baseline to determine W$_{10}$. 

\subsection*{PSR\,J1643$+$1338 -- anti-RFM profile evolution}

PSR\,J1643$+$1338 shows two distinct peaks in its profiles at all three observing frequencies, all of which are well modelled with just a single component per peak. Both W$_{10}$ and W$_{50}$ of the pulsar are shown to increase with increasing frequency. We also measured the separation between two components (Table~\ref{tab:componentsep}). We found that the separation between the components increases at higher frequencies as well, suggesting that the pulsar does not conform to the RFM model, where the component separations are expected to decrease with increasing frequencies. While the behaviour of PSR\,J1643$+$1338 is unusual, the anti-RFM profile evolution has been seen in several other pulsars before~\citep{hsh+12,cw14,nsk+15,phs+16}.

\begin{table}
\centering
\caption[The separations between the two major components in PSRs\,J1643$+$1338, J1749$+$5952 and J1916$+$3224 at 149, 334 and 1532\,MHz]{The separation between the 2 major pulse profile components in PSRs\,J1643$+$1338, J1749$+$5952 and J1916$+$3224 at 149, 334 and 1532\,MHz.}
\begin{tabular}{lcc}
\hline
Pulsar & Frequency & Separation between components\\
PSR & MHz & ms \\
\hline
J1643$+$1338 & 149 & 10.1(1)\\
 & 334 & 12.6(5)\\
 & 1532 & 14.5(3)\\
J1749$+$5952 & 149 & 11.35(8)\\
 & 334 & 10.0(5)\\
 & 1532 & 8.4(2)\\
J1916$+$3224 & 149 & 15.6(5)\\
 & 334 & 12(2)\\
 & 1532 & 10.9(12)\\
 \hline
\end{tabular}
\label{tab:componentsep}
\end{table}

\subsection*{PSR\,J1749$+$5952 -- narrowing profile components at lower frequencies}

The pulse profiles of PSR\,J1749$+$5952 show two distinct peaks at 149 and 334\,MHz, while at 1532\,MHz, the trailing peak seems to merge with the leading peak. The profiles at 334 and 1532\,MHz can be modelled relatively well with just two components. However, the profile at 149\,MHz requires three components; the two main components describing each peak and a bridge component that is fitted across the profile. We measured two distinct sets of W$_{50}$ values of the profiles at both 149 and 334\,MHz. We found that this is due to the trailing peak of both profiles having roughly half the intensity of the leading peak. The smaller values in Table~\ref{tab:widths} correspond to the width of the main peak at half maximum while the larger values correspond to the width of the whole profile.

The measurements of the separation of the components corresponding to the two peaks seen in the profile of J1749$+$5952, shown in Table~\ref{tab:componentsep} suggest that the separation between components decreases with increasing frequency, in agreement with RFM. However, the values of W$_{50}$ that correspond to the width of the main peak at 149 and 334\,MHz suggest that the width of the main peak increases with increasing frequency. The increase in width of the main peak at higher frequency could be affected by the exact beam pattern of the pulsar, and pulse longitude dependent spectral index effects. 

\subsection*{PSR\,J1810$+$0705 -- large profile evolution}

The integrated pulse profiles of PSR\,J1810$+$0705 show significant frequency evolution. At 149\,MHz, the profile shows two distinct peaks of about equal intensity and at 1532\,MHz, the intensity of the leading peak relative to the trailing peak is reduced. The pulsar shows two distinct peaks at 334\,MHz as well, however, the low S/N of the observation does not allow us to get a good estimate of the relative intensity between the peaks. The W$_{10}$ of the pulsar also shows a large decrease with increasing frequency, suggesting that the separation between the two components decreases with increasing frequency as expected. Due to the complex structure of the profile of PSR\,J1810$+$0705 at 149\,MHz, we are unable to model the two peaks as distinct components to measure the component separation. The templates that were used for pulsar timing requires several components for each peak to produce an adequate model of the pulse profiles.

\subsection*{PSR\,J1916$+$3224 -- standard profile evolution}

PSR\,J1916$+$3224 shows a strong main peak and a trailing, wider component at all three observing frequencies. The evolution of the pulse profile is found to follow the RFM model, where the W$_{50}$ of the profiles, which correspond to the width of the main peaks, decreases as observing frequency increases. While the measured W$_{10}$ suggests the profile width is constant between 334 and 1532\,MHz, the separation between the two main components (Table~\ref{tab:componentsep}) decreases with increasing observing frequency, in agreement with RFM model.

\section{Flux densities and spectral indices measurements} \label{sec:spectra}

The flux densities and the measured spectral indices of the isolated pulsars except PSR\,J1657$+$3304 are shown in Table~\ref{tab:flux}. Only pulsars detected by LOFAR and another observing frequency with the Lovell Telescope have a measured spectral index. An upper limit on the spectral index is given for pulsars with measured flux densities only in the LOFAR band. The spectral indices of the pulsars are modelled using a single power law where $S_{\nu} \propto \nu^{\alpha}$, where $S$ is the flux density, $\nu$ is the observing frequency and $\alpha$ is the spectral index of the pulsar.

\begin{table*}
\centering
\caption{Flux density measurements and modelled spectral indices of the 20 pulsars described in Section~\ref{sec:timingprop}.}
\label{tab:flux}
\begin{tabular}{lcccccc}
\hline
 PSR & & \multicolumn{4}{c}{$\nu$ (MHz)} & $\alpha$\\
 & & 128 & 167 & 334 & 1532 & \\
\hline
J0100$+$8023 & & 2.4(7) & 1.5(4) & $<$0.3 & $<$0.03 & $<-$2.2\\
J0107$+$1322 & & 2.8(10) & 1.8(7) & 0.46(9) & 0.07(3) & $-$1.5(5)\\
J0210$+$5845$^{a}$ & & 3.4(8) & 3.1(6) & 0.8(2) & $<$0.05 & $-$1.6(8)$^{d}$\\
J0421$+$3255 & & 27(6) & 16(4) & $<$2.0 & $<$0.16 & $<-$2.8\\
J0454$+$4529$^{a}$ & & 7.0(17) & 2.7(7) & 0.44(9) & $<$0.04 & $-$2.8(7)\\
J1017$+$3011$^{a}$ & & 3.4(5) & 1.9(4) & 0.31(6) & $<$0.06 & $-$2.5(6)\\
J1624$+$5850 & & 2.5(5) & 1.3(3) & $<$0.4 & $<$0.04 & $<-$1.9\\
J1638$+$4005 & & 3.1(5) & 1.7(3) & 0.34(7) & $<$0.06 & $-$2.3(6)\\
J1643$+$1338 & & 11(4) & 9(3) & 2.6(5) & 0.11(3) & $-$1.9(4)\\
J1656$+$6203 & & 2.6(5) & 1.2(3) & $<$0.5 & $<$0.04 & $<-$1.6\\
J1657$+$3304$^{b}$ & & -- & -- & -- & -- & --\\
J1713$+$7810 & & 1.7(6) & 0.9(3) & $<$0.6 & $<$0.07 & $<-$1.2\\
J1741$+$3855 & & 3.1(7) & 2.0(5) & $<$0.5 & $<$0.06 & $<-$1.9\\
J1745$+$1252 & & 9.6(16) & 5.5(15) & 2.2(4) & $<$0.08 & $-$1.5(6)$^{d}$\\
J1749$+$5952 & & 5.9(21) & 4.1(14) & 2.3(5) & 0.16(7) & $-$1.4(5)\\
J1810$+$0705 & & 14(4) & 11(3) & 2.6(5) & 0.22(6) & $-$1.7(3)\\
J1916$+$3224 & & 3.0(7) & 2.1(6) & 1.7(3) & 0.08(3) & $-$1.3(4)$^{d}$\\
J1957$-$0002 & & 7.5(15) & 3.5(7) & $<$0.5 & $<$0.03 & $<-$2.8\\
J2034$+$6646$^{a}$ & & 5.0(10) & 2.1(6) & --$^{c}$ & $<$0.06 & $<-$1.7\\
J2122$+$2426 & & 4.0(12) & 2.0(9) & --$^{c}$ & 0.07(2) & $<-$1.6\\
\hline
\multicolumn{7}{p{9cm}}{$^{a}$These pulsars have no observations that were pointed towards their timing positions, which might result in smaller measured flux.}\\
\multicolumn{7}{p{9cm}}{$^{b}$PSR\,J1657$+$3304 showed large flux density variation across the observing span with LOFAR (see Section~\ref{sec:J1657}).}\\
\multicolumn{7}{p{9cm}}{$^{c}$The observations are strongly affected by RFI. Hence, we were not able to obtain a reliable flux density estimates from them.}\\
\multicolumn{7}{p{9cm}}{$^{d}$The spectra of these pulsars are not well-described by a single power law.}\\
\end{tabular}
\end{table*}

We then compute the average spectral index of the 19 pulsars, while using the upper limits as the measured spectral index whenever it is relevant. We found the average spectral index of the 19 LOTAAS discoveries studied here to be $\bar{\alpha} = -1.9 \pm 0.5$. The average is similar to $\bar{\alpha} = -1.8$ found by~\citet{mkkw00}, but slightly steeper than the average of $\bar{\alpha} = -1.4$ found by~\citet{blv13} and~\citet{bkk+16}, as well as $\bar{\alpha} = -1.6$ by~\citet{jsk+18}, respectively. As the spectral indices of 9 out of 19 pulsars are upper limits, the actual average spectral index of these pulsars can be even lower. This suggest that the sample of pulsars studied here have relatively steep spectra, and that LOTAAS is discovering steep spectra pulsars that are otherwise not detectable with surveys at higher frequencies.

\begin{figure}
  \centering
  \includegraphics[width=0.9\linewidth]{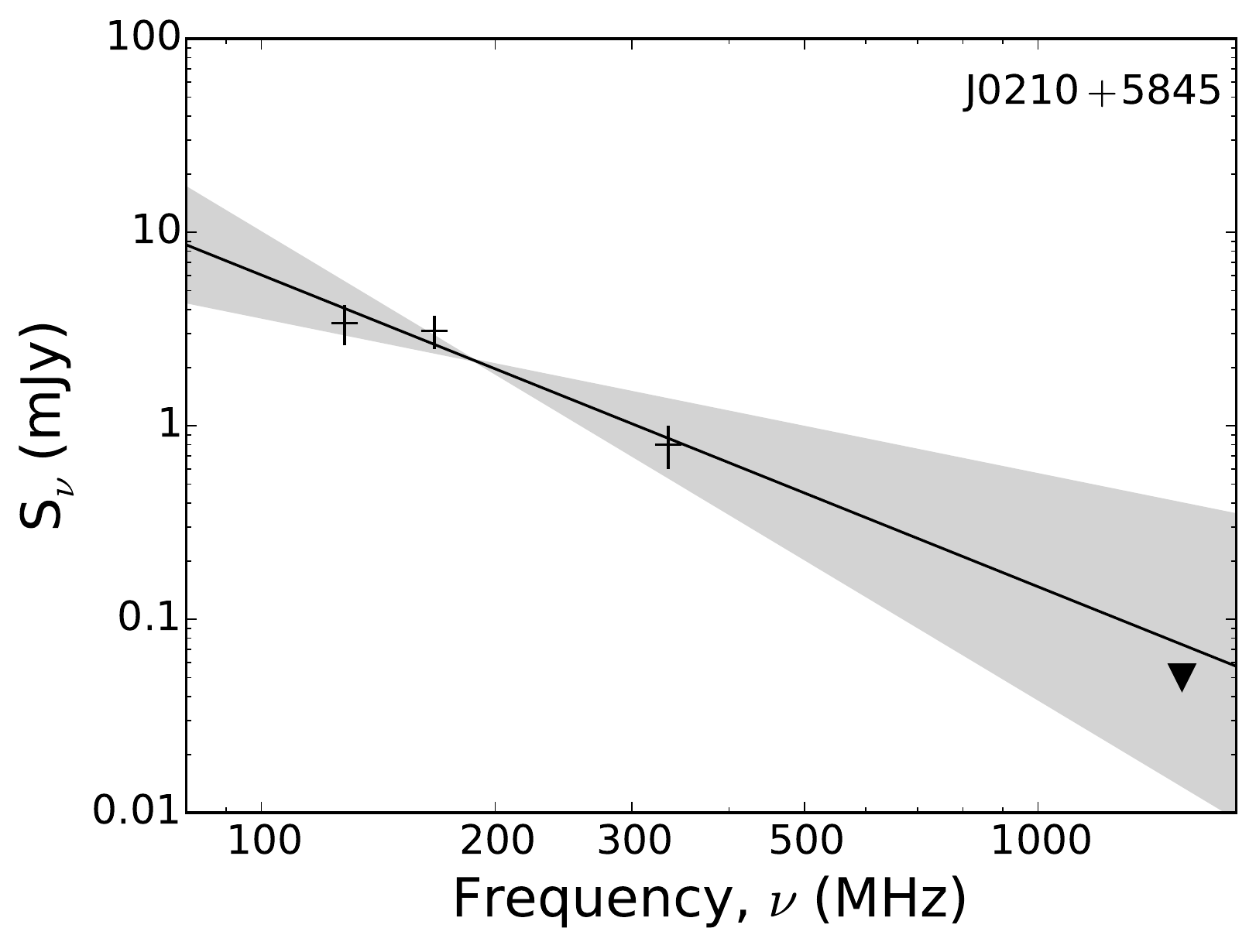}
  \includegraphics[width=0.9\linewidth]{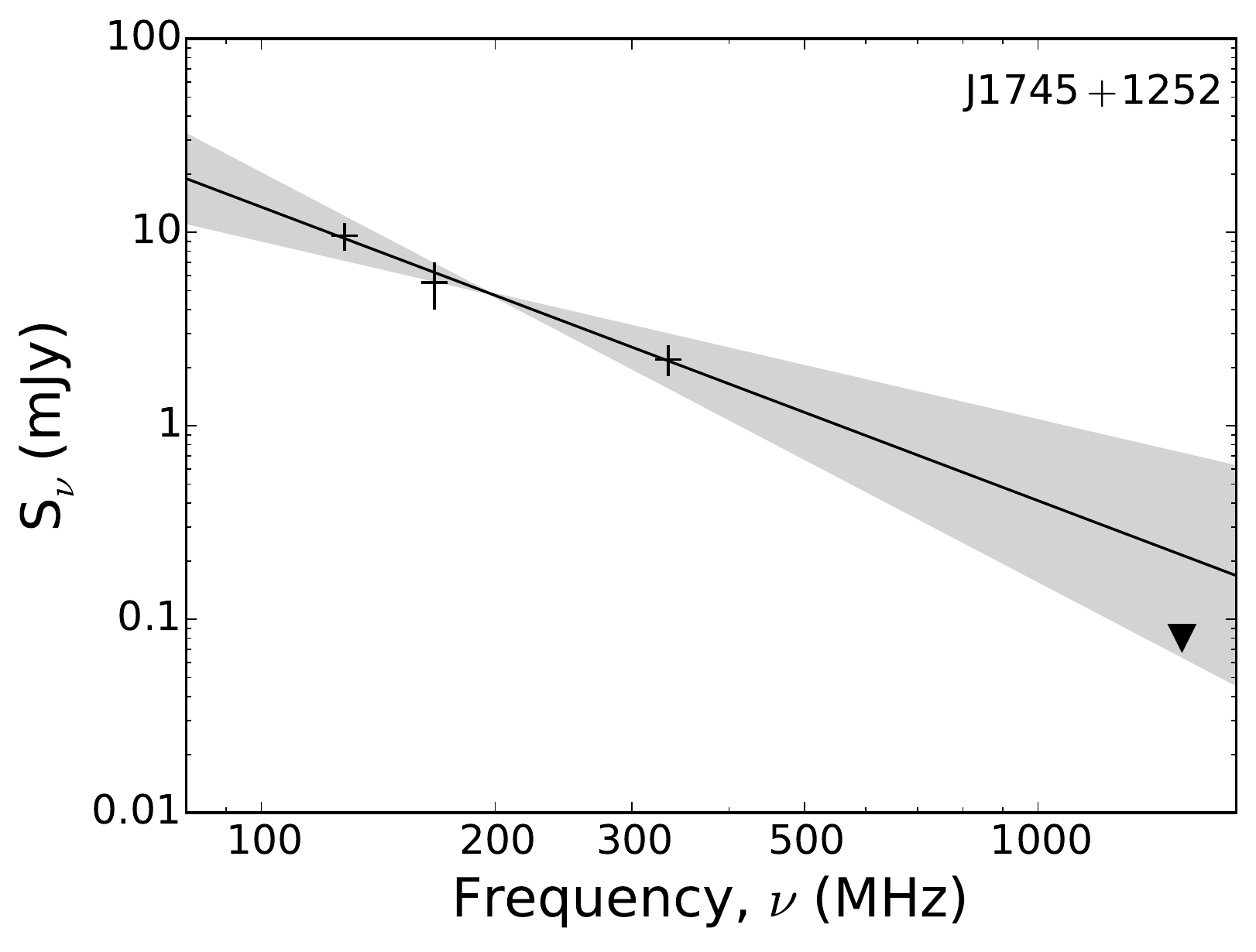}
  \includegraphics[width=0.9\linewidth]{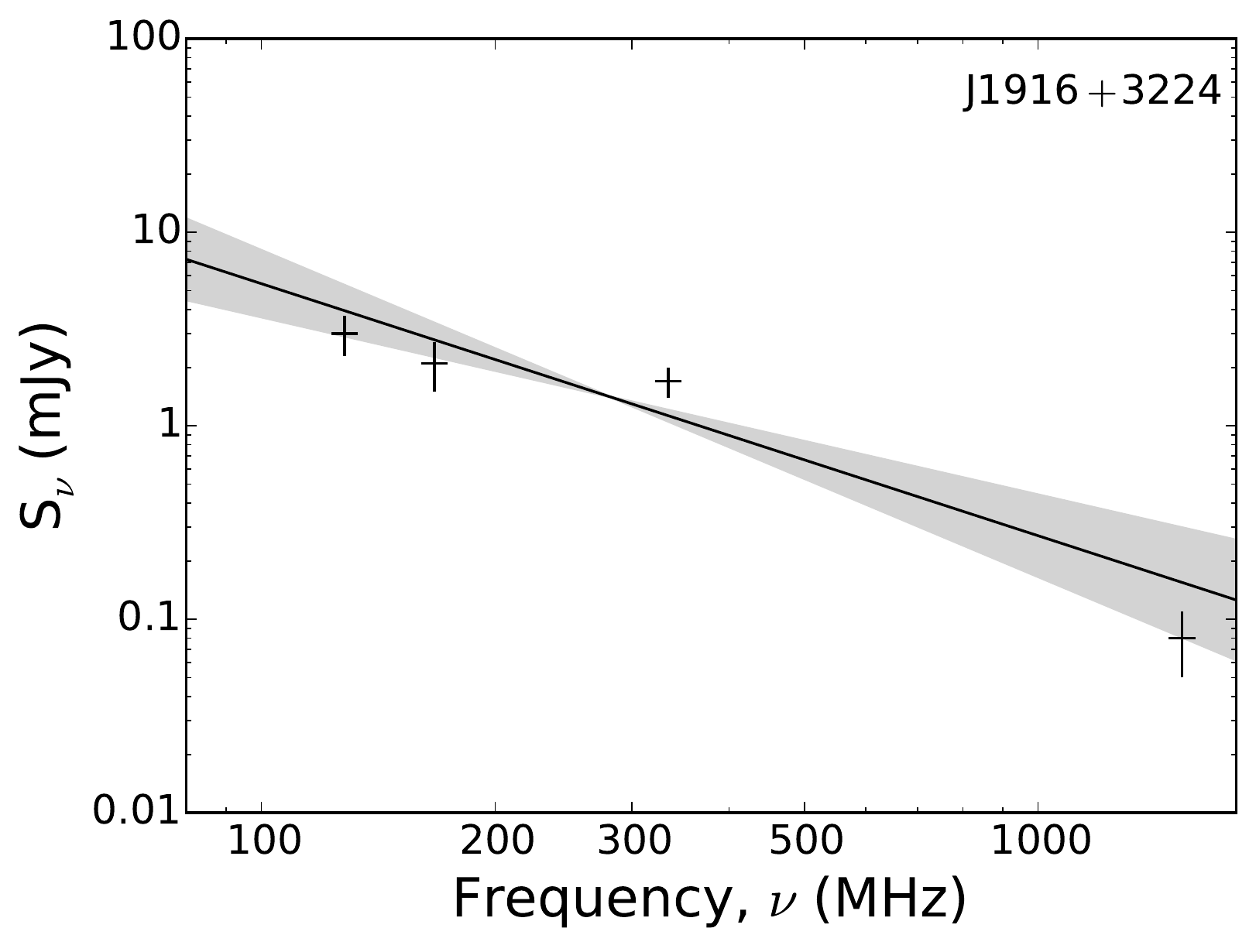}
\caption[The spectra of PSRs\,J0210$+$5845, J1745$+$1252 and J1916$+$3224.]{The spectra of PSRs\,J0210$+$5845 (\textit{top}), J1745$+$1252 (\textit{centre}) and J1916$+$3224 (\textit{bottom}), fitted with a simple power law. The fitted spectral indices are shown as black lines, with a 1$\sigma$ uncertainty region shaded in grey. The triangles denote the upper limits on flux densities from the non-detections at the relevant observing frequencies, assuming a minimum S/N detection threshold of 10. The fitted spectral indices and uncertainties can be found in Table~\ref{tab:flux}.}
\label{fig:badspectralfit}
\end{figure}

The spectra of three of the pulsars are not well-described by a single power law. Fig.~\ref{fig:badspectralfit} shows the spectral fit of PSRs\,J0210$+$5845, J1745$+$1252 and J1916$+$3224 with a single power law. The flux densities of PSR\,J0210$+$5845 suggest that the spectrum is less steep at 128 and 167\,MHz and steeper at higher frequencies, indicating that the spectrum could be a broken power law with two different spectral indices below and above a critical frequency between 167 and 334\,MHz, with the spectrum becoming steeper at higher frequencies. The steepening of pulsar spectra at high frequencies has been observed in many other pulsars before~\citep[e.g.][]{sie73,mkkw00,bkk+16}. There is also a possibility that a spectral turnover occured at the LOFAR observing frequencies of about 150\,MHz, which would require observations at even lower frequencies to confirm. The spectrum of PSR\,J1745$+$1252 is well fit by a single power law between 128 and 334\,MHz. However, the upper limit on flux density at 1532\,MHz suggests that the spectrum becomes steeper at higher observing frequencies. The spectrum of PSR\,J1916$+$3224 suggests that the flux density at 334\,MHz is too large and too small at 1532\,MHz when fitted with a single power law. The spectrum is more likely to be a broken power law with a more negative spectral index at higher frequencies, possibly above 334\,MHz.

The fraction of pulsars studied here that were not well-fit with a single power law is in line with what is found by~\citet{bkk+16} and~\citet{jsk+18}. In those cases, the pulsars are either fitted with a broken power law spectrum or a log-parabolic spectrum with a turnover frequency. The three pulsars discussed are unlikely to be part of the Gigahertz-peaked spectrum pulsars~\citep{klm+11,rla16}, as the flux densities measured at 334\,MHz are lower than the flux densities at LOFAR frequencies. The Gigahertz-peaked spectrum pulsars are expected to have turnover frequencies of between 0.6$-$2\,GHz. In order to better study the spectra of these sources and other LOTAAS discoveries, we would require observations conducted at frequencies not covered by this work.

\section{Individual Pulsars} \label{sec:interestingpulsars}

\subsection{PSR\,J1657$+$3304 -- large flux density variation, nulling and mode changing} \label{sec:J1657}

\begin{figure}
  \centering
  \includegraphics[width=\linewidth]{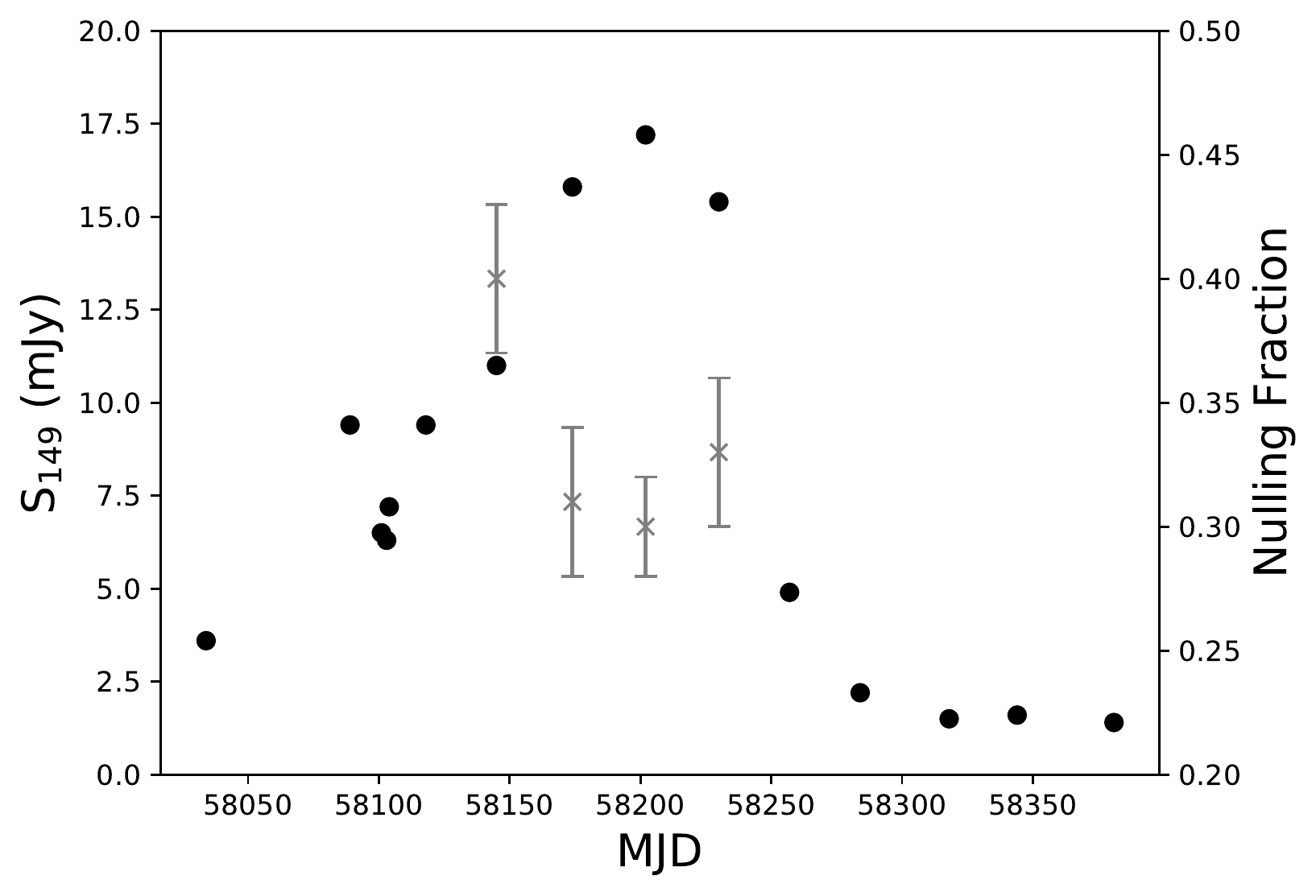}
  \caption{Plot showing the flux density at 149\,MHz (black dots) and the estimated nulling fraction (grey crosses), with 1$\sigma$ uncertainty, of individual observations of PSR\,J1657$+$3304 taken with LOFAR. The nulling fractions are only estimated for observations with measured flux density above 10 mJy due to the unreliability of the estimates from observations with lower measured flux densities. The uncertainty of individual flux density measurements is estimated to be 50 per cent.}
  \label{fig:J1657fluxnulling}
\end{figure}

The emission of PSR\,J1657$+$3304 shows several interesting properties. Firstly, it shows long-term flux density variation over the observation span. We measured its flux density at 149\,MHz across 15 observations over 347 days, shown in Fig.~\ref{fig:J1657fluxnulling}, and it varies by a factor of 10 over approximately 300 days before staying constant at low flux density (1.5\,mJy) after MJD 58300. The magnitude of the variation is much larger than any other pulsars discussed in this work. We attempted to identify any potential period derivative change that correlates with the flux density variation, as the changes in flux density of the pulsar could be due to long-term mode changing that is often accompanied with changes in the period derivative of the pulsar~\citep{klb+06,lhk+10}. We did not detect any noticeable change in period derivative.


\begin{figure*}
\centering
\includegraphics[width=0.49\linewidth]{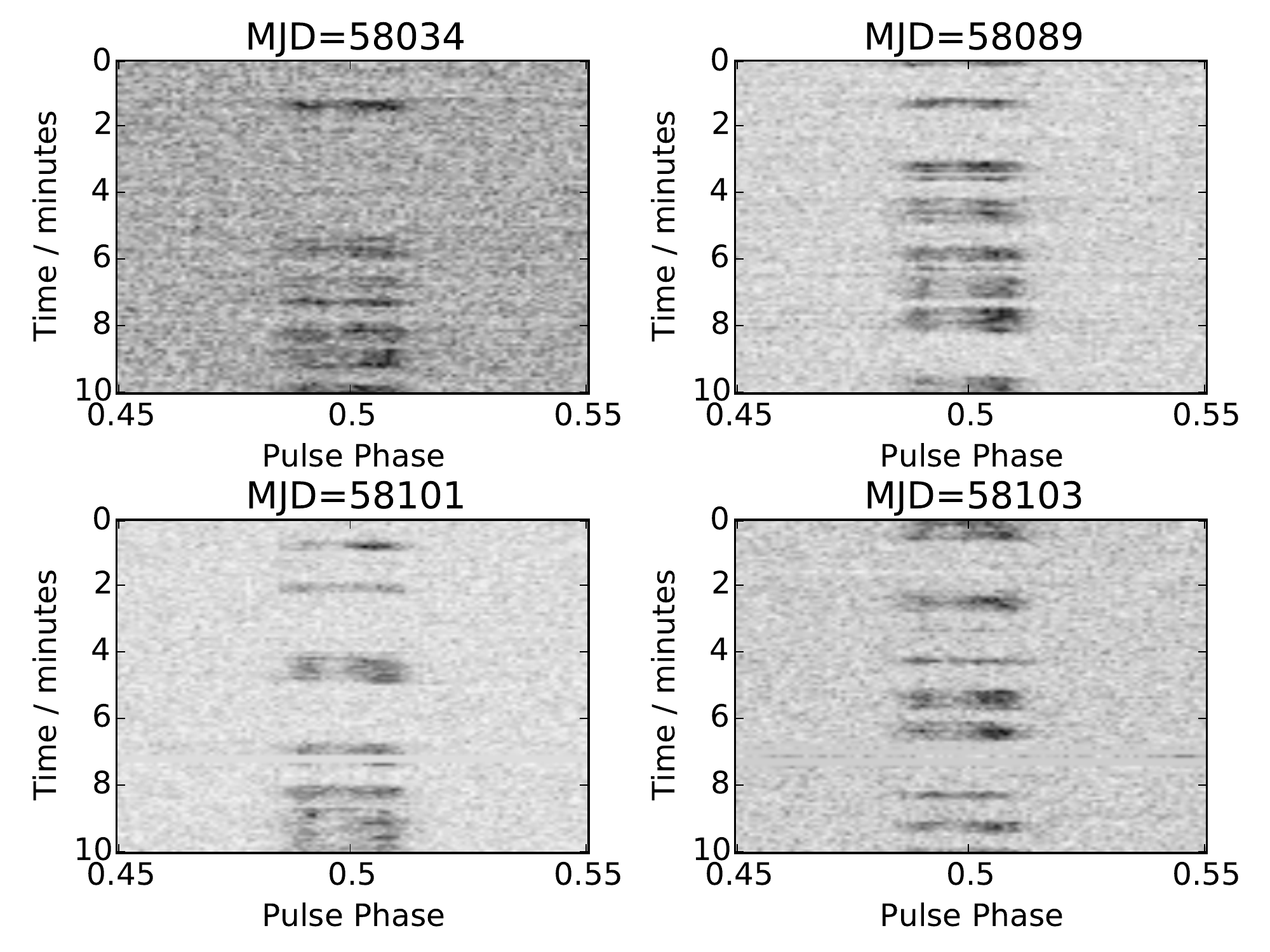}
\includegraphics[width=0.49\linewidth]{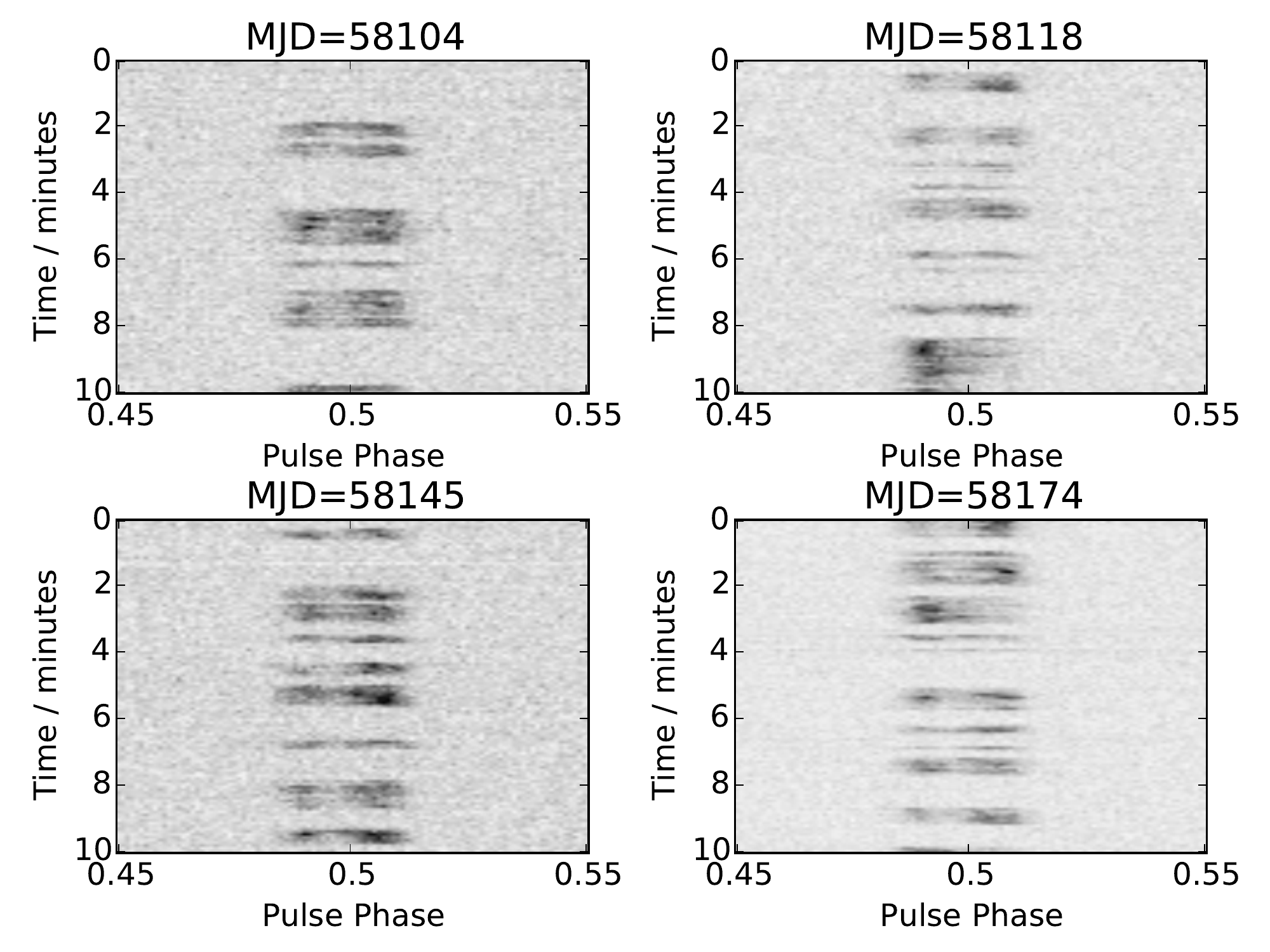}
\includegraphics[width=0.49\linewidth]{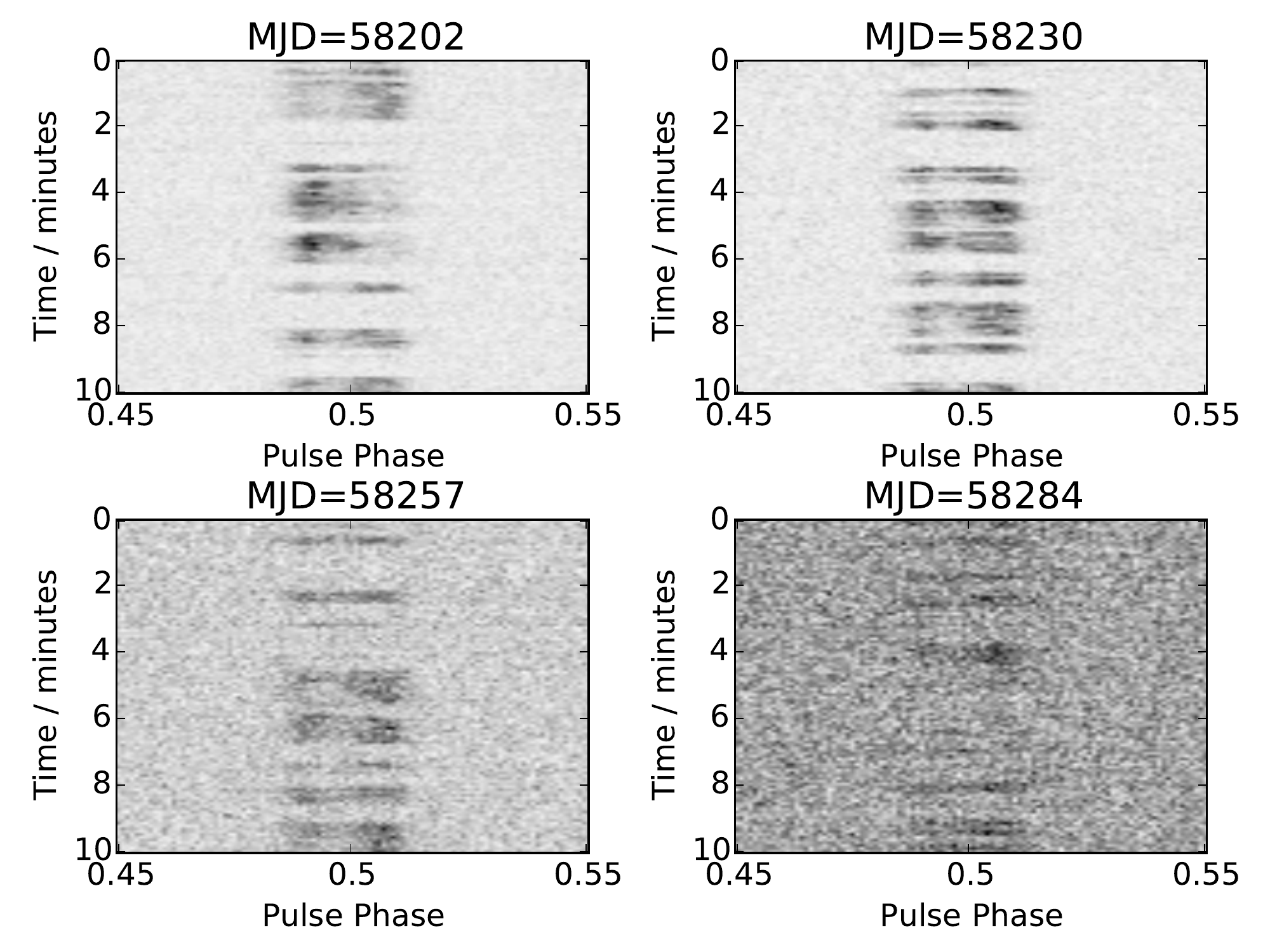}
\includegraphics[width=0.49\linewidth]{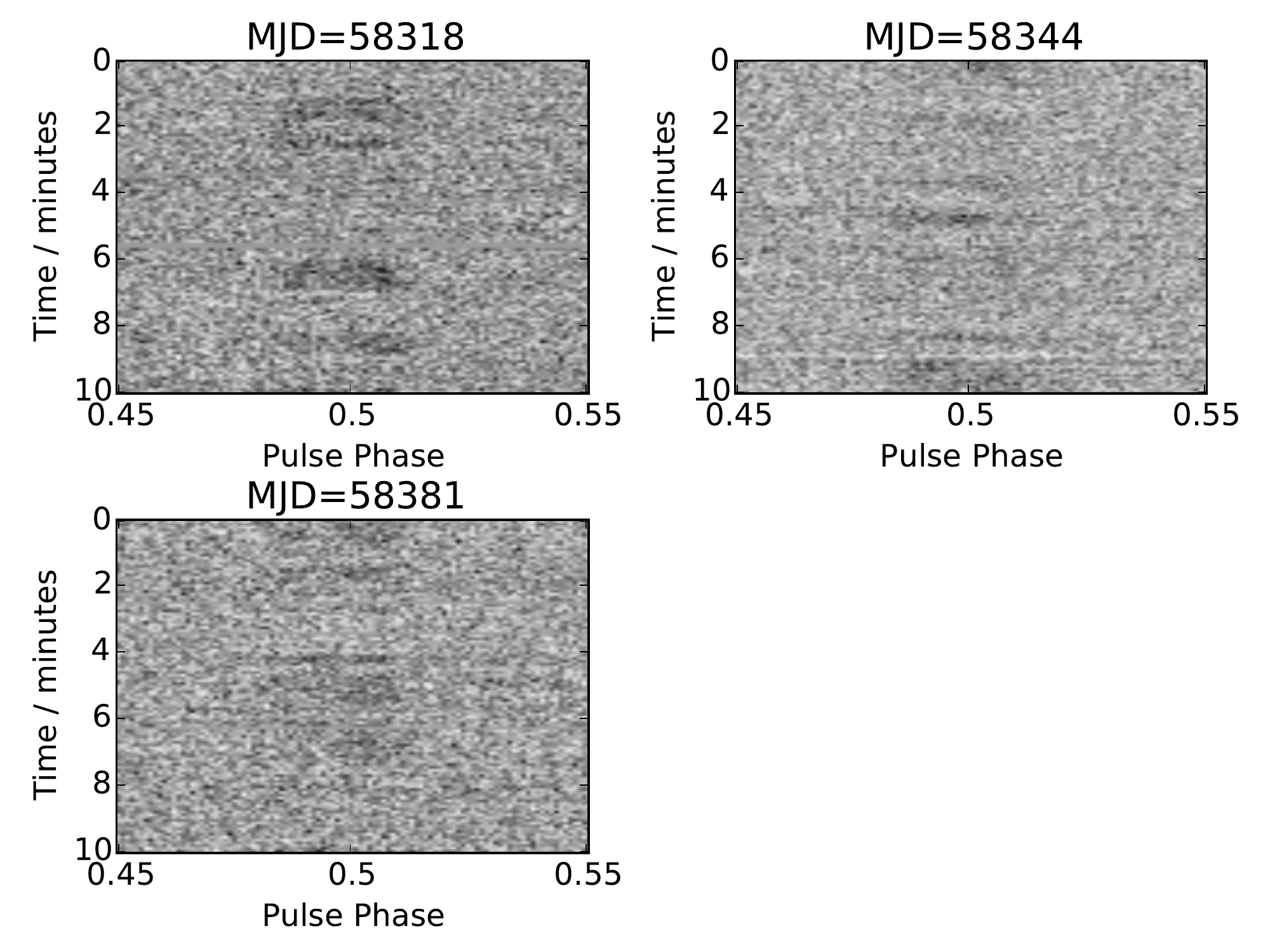}
\caption[The sub-integration against pulse phase plots of 15 different observations of PSR\,J1657$+$3304.]{The sub-integration against pulse phase plots of all 15 different observations of PSR\,J1657$+$3304 presented in Fig.~\ref{fig:J1657fluxnulling}. Each sub-integration is 5\,s long, corresponding to between 3-4 pulses of the pulsar. There is a trend of increasing flux density of the pulsar from MJD 58034 to about MJD 58100 and decreasing flux density from MJD 58230 to MJD 58284. The nulling of the pulsar can also be seen in most of the observations; the shortest null is about 5\,s as seen in the observation taken on MJD 58230 and the longest null seen in these observations is about 4\,minutes from the observations taken on MJD 58034 and MJD 58284. The observation taken on MJD 58202 shows clear mode changing between the on-phase at 0-2 minutes and at 3-6 minutes, where the profile switches from a double peaked structure to a single peaked structure. The plots are zoomed into 10\% of the pulse phase of the pulsar.}
\label{fig:J1657timeplots}
\end{figure*}

We also looked into the possibility that the flux density variation is due to interstellar scintillation in the strong regime. We found that the variation is unlikely to be due to diffractive interstellar scintillation as the expected scintillation bandwidth at 149\,MHz is less than 1\,kHz according to the NE2001 model~\citep{cl02}, much smaller than the bandwidth of the LOFAR timing observations of 78\,MHz. The long-term flux density variations of pulsars are often due to refractive interstellar scintillation~\citep[RISS;][]{rcb84}. However, we found that this is unlikely for PSR\,J1657$+$3304, as the expected timescale of RISS, $t_\textrm{RISS}$ has a frequency dependence of $t_\textrm{RISS} \propto \nu^{-2.2}$, which means that the timescale in the changes in flux density of the pulsar between the top (167\,MHz) and bottom half (128\,MHz) of the LOFAR band is expected to be different by a factor of 1.8. We found that the changes in flux density of the pulsar to be consistent across the LOFAR bandwidth. Hence, it is more likely that the flux density variation of PSR\,J1657$+$3304 is intrinsic to the pulsar.

Individual observations of the pulsar also showed nulling over duration between several pulses to a few minutes (Fig.~\ref{fig:J1657timeplots}). We estimated the nulling fraction of each observation of PSR\,J1657$+$3304 following the method of~\citet{wmj07}, where the pulse energy distribution of the pulsar is compared with the off-pulse energy distribution. We also tested whether the obtained nulling fractions are the same for both high and low flux density observations, by increasing the noise level of the high flux density observations to have the same S/N as the low flux density observations and then calculating the nulling fraction. We find that the measured nulling fraction changes with S/N as it is more difficult to separate the pulses from the nulls. Hence we only estimated the nulling fraction from the four observations with flux densities above 10 mJy. The average nulling fraction is found to be 34 per cent. While the presence of nulling could account for some of the flux density variation seen between observations, we found that the flux density of individual sub-integrations also varied between observations. 

\begin{figure}
\centering
\includegraphics[width=\linewidth]{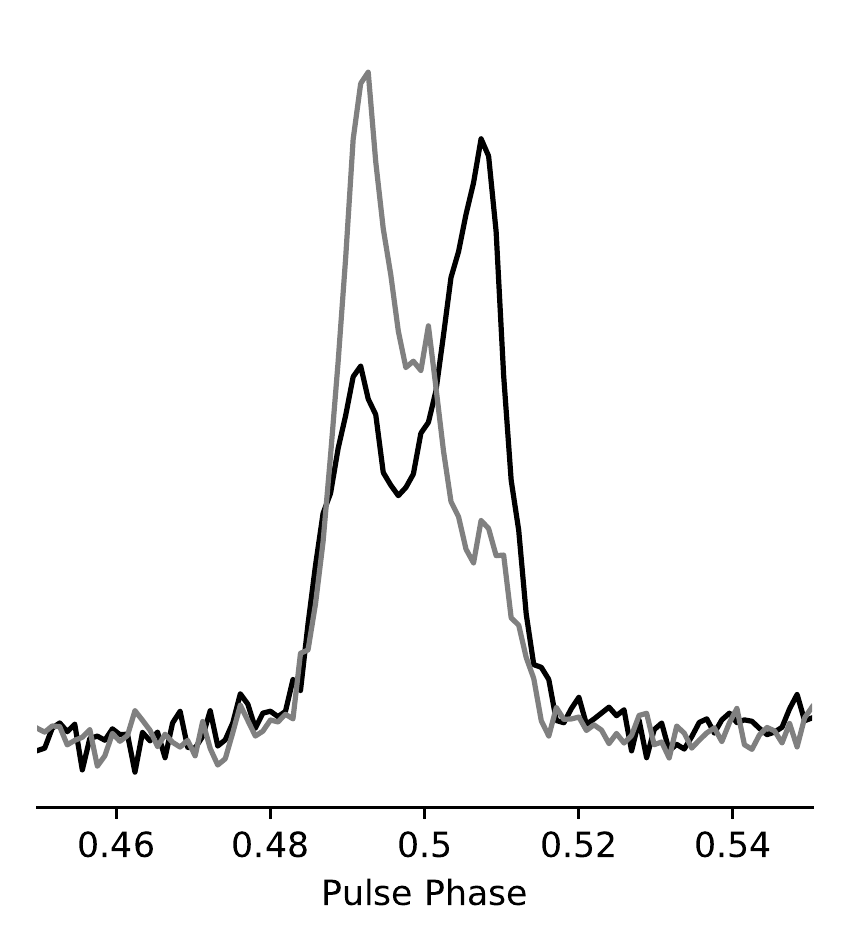}
\caption[The integrated pulse profiles of the two different modes of PSR\,J1657$+$3304]{The integrated pulse profiles of the two different modes of PSR\,J1657$+$3304, overlaid on top of each other. The more commonly occurring double-peaked mode is in black and the less common single-peaked mode is in grey. The integrated pulse profiles are normalized via the off-pulse standard deviation.}
\label{fig:J1657modes}
\end{figure}

In addition to nulling, we observed short-term mode changing in some observations, where the pulse profile switches between the commonly occurring double-peaked structure to only the leading peak being present (Fig.~\ref{fig:J1657modes}). The duration of the occurrence of each instance of the less common mode ranged between 1-5 minutes, similar to the duration of the nulls. For example, the less common mode occurred between third and sixth minute of the observation in MJD 58202, as shown in Fig.~\ref{fig:J1657timeplots}. Most of the mode changing occurred between nulls, with several exceptions such as at the beginning of the observation in MJD 58118 where the pulsar changes mode before a period of nulling.

PSR\,J1657$+$3304 was observed with the Lovell Telescope at both 334 and 1532\,MHz. The observation at 334\,MHz is strongly affected by RFI, and the pulsar is not detected in four 30-minute observations at 1532\,MHz. However, the observations at 1532\,MHz coincide with the period when PSR\,J1657$+$3304 showed low flux density at LOFAR observing frequencies. This gives an upper limit to the flux density of the pulsar at 1532\,MHz of 0.06\,mJy when it is in the low flux density state.

\subsection{PSR\,J1658$+$3630 -- binary pulsar} \label{sec:binarypsr}

\begin{figure} 
\includegraphics[width=\linewidth]{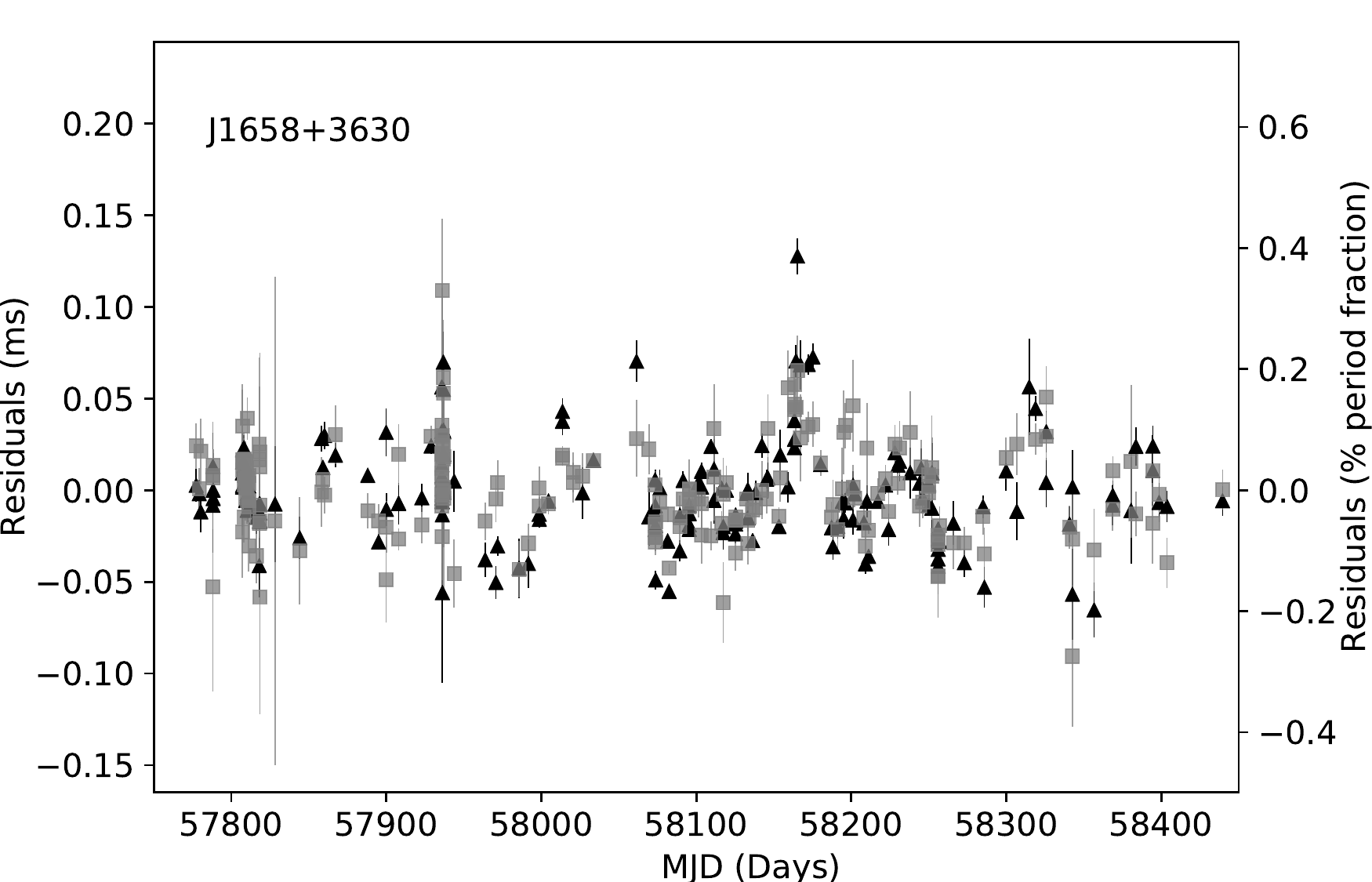}
\includegraphics[width=\linewidth]{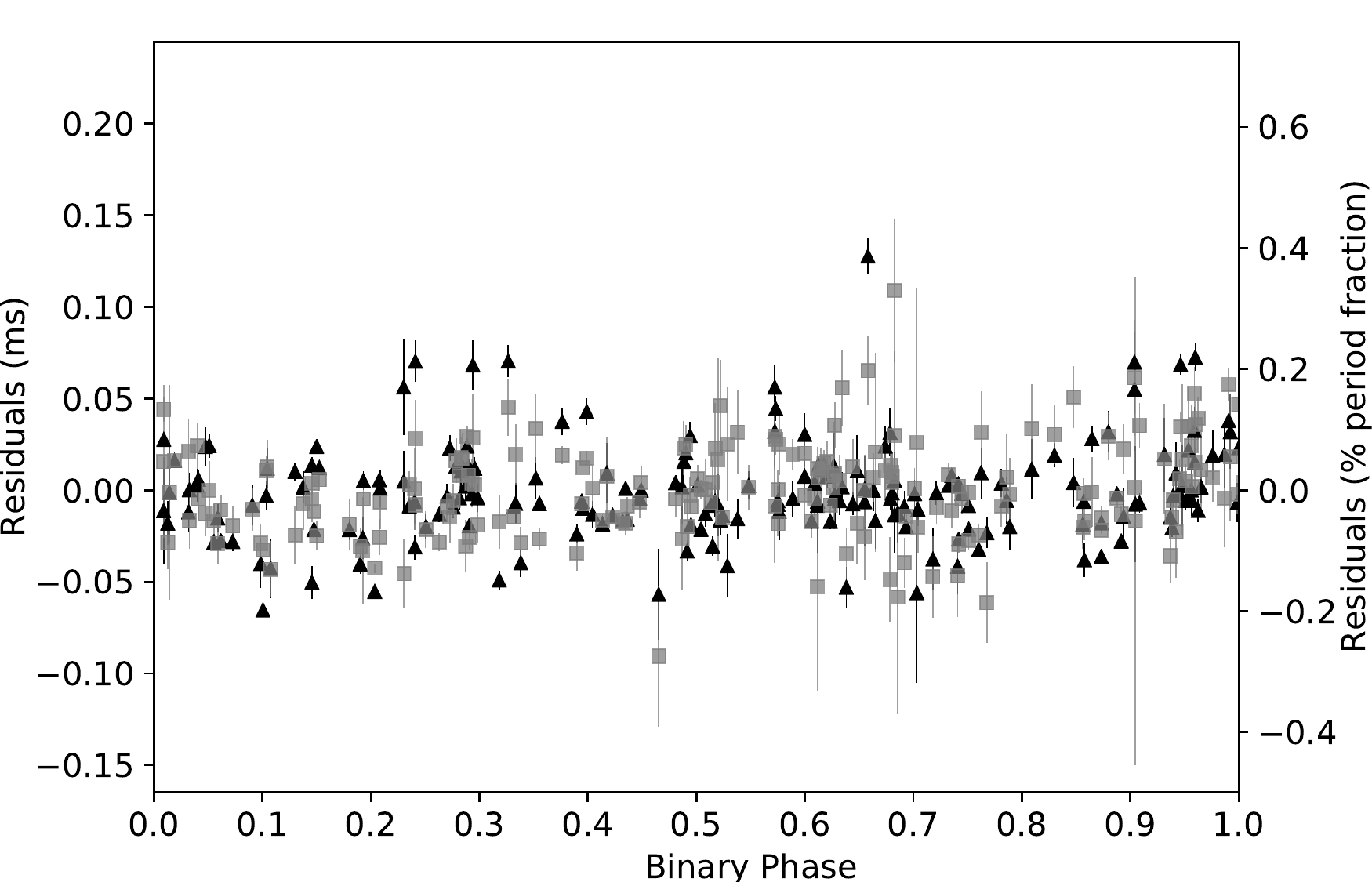}
\caption{The timing residuals of PSR\,J1658$+$3630 over time (\textit{top}) and in terms of the binary orbital phase of the pulsar (\textit{bottom}). The different symbols represents the different observing frequencies, with the black triangles representing the lower part of the LOFAR and GLOW bands (central frequencies of 128\,MHz and 140\,MHz, respectively) and grey squares for the upper part of the LOFAR and GLOW bands (central frequencies of 167\,MHz and 171\,MHz, respectively). The timing residuals over time suggest that there might be unmodelled DM variation over timescale shorter than 30 days, while the timing residuals across the binary phase do not show  variation that can be attributed to Shapiro delay due to the binary companion.}
\label{fig:1658residual}
\end{figure}

The low eccentricity of the orbit of PSR\,J1658$+$3630 led us to use the ELL1 binary timing model~\citep{lcw+01} to model the TOAs. The parameters of the timing model are shown in Table~\ref{tab:J1658timing} and the timing residuals as a function of both time and binary phase are shown in Fig.~\ref{fig:1658residual}. The pulsar has a spin period of 33\,ms and spin period derivative of 1.16$\times 10^{-19}$ s s$^{-1}$, placing it in the location of the $P$-$\dot{P}$ diagram (Fig.~\ref{fig:LOTAAStimingppdot}) that is populated by the intermediate-mass binary pulsars (IMBP), suggesting that PSR\,J1658$+$3630 is part of the IMBP population. The average DM of the pulsar is 3.044 pc cm$^{-3}$, indicating DM-distances of either 0.22~\citep{ymw17} or 0.49 kpc~\citep{cl02}, making PSR\,J1658$+$3630 one of the closest-known pulsar binaries. The orbital period is 3.016 days and the companion has a minimum mass of 0.87 $M_{\odot}$. The minimum mass of the companion is sufficiently high that the measurement of Shapiro delay is possible if the system has a high inclination angle. However, we found that the pulsar showed temporal variation in DM, which has a large effect on the TOAs at LOFAR frequencies, we are unable to measure any Shapiro delay, and that the timing residuals in terms of binary orbital phase (Fig.~\ref{fig:1658residual}) does not show signs of Shapiro delay (See \citealt{dpr+10} for the effects of Shapiro delay on the timing residuals).

\begin{table*}
\centering
\caption[Timing solution for PSR\,J1658$+$3630.]{Timing Solution of PSR\,J1658$+$3630, obtained by fitting the TOAs with the rotational, positional and orbital parameters of the pulsar. The proper motion of the pulsar is fixed at the estimated value obtained through archival optical imaging data.}
\begin{tabular}{lcc}
\hline
Timing Parameter & Value\\
\hline
Right Ascension, RA (J2000) & 16:58:26.5198(3)\\
Declination, Dec (J2000) & $+$36:30:30.095(3)\\
Spin frequency, $\nu$ (s$^{-1}$) & 30.277356639727(13)\\
Spin frequency Derivative, $\dot{\nu}$ (10$^{-16}$ s$^-2$) & $-1.061$(17)\\
DM$^a$ (pc cm$^{-3}$) & 3.04387(3)\\
Proper motion, RA (arcsec yr$^{-1}$) & 0.015(8)\\
Proper motion, Dec (arcsec yr$^{-1}$) & $-$0.065(8)\\
Epoch of timing solution (MJD) & 58073\\
Epoch of position (MJD) & 58073\\
Solar system ephemeris model & DE405\\
Clock correction procedure & TT(TAI)\\
Time units & TCB\\
Timing Span (MJD) & $57777-58439$\\
Number of TOAs & 399\\
Weighted post-fit residual ($\upmu$s) & 20.6\\
Reduced $\chi^2$ value & 5.3\\
\hline
Binary Parameter & \\
\hline
Orbital period, $P_\mathrm{b}$ (days) & 3.0163073825(13) \\
Projected Semi-major Axis, $a_{\mathrm{p}} \mathrm{sin}\,i$ (lt-s) & 10.4497253(14) \\
Epoch of Ascending Node, $T_\mathrm{asc}$ (MJD) & 57768.17408579(15)\\
First Laplace-Lagrange parameter, $\epsilon_1$ (10$^{-5}$) & 2.14(3)\\
Second Laplace-Lagrange parameter, $\epsilon_2$ (10$^{-5}$) & $-$2.17(3)\\
Binary Model & ELL1\\
\hline
Derived Parameter & \\
\hline
Spin Period, $P$ (s) & 0.033027982326829(13)\\
Spin Period Derivative, $\dot{P}$ (10$^{-20}$ s s$^{-1}$) & 11.57(19)\\
Orbital Eccentricity, $e$ (10$^{-5}$) & 3.05(3)\\
Longitude of Periastron, $\omega$ ($\degr$) & 135.3(5)\\
Epoch of Periastron, $T_0$ (MJD) & 57769.308(4)\\
$\mathrm{DM}$ distance/NE2001 (kpc) & 0.49\\
$\mathrm{DM}$ distance/YMW16 (kpc) & 0.22\\
Characteristic Age (Gyr) & 4.5\\
Surface Magnetic Field Strength (10$^9$ G) & 2.0\\
Mass Function (M$_{\odot}$) & 0.13466234(2)\\
Minimum Companion Mass (M$_{\odot}$) & 0.8731\\
Spin-down luminosity (10$^{30}$ erg s$^{-1}$) & 127\\
Spectral index, $\alpha$ & $-$2.5(7)\\
\hline
\multicolumn{3}{p{9.5cm}}{$^{a}$This is the reference DM to measure the temporal DM variation of the pulsar.}\\
\end{tabular}
\label{tab:J1658timing}
\end{table*}

\begin{figure}
\centering
\includegraphics[width=\linewidth]{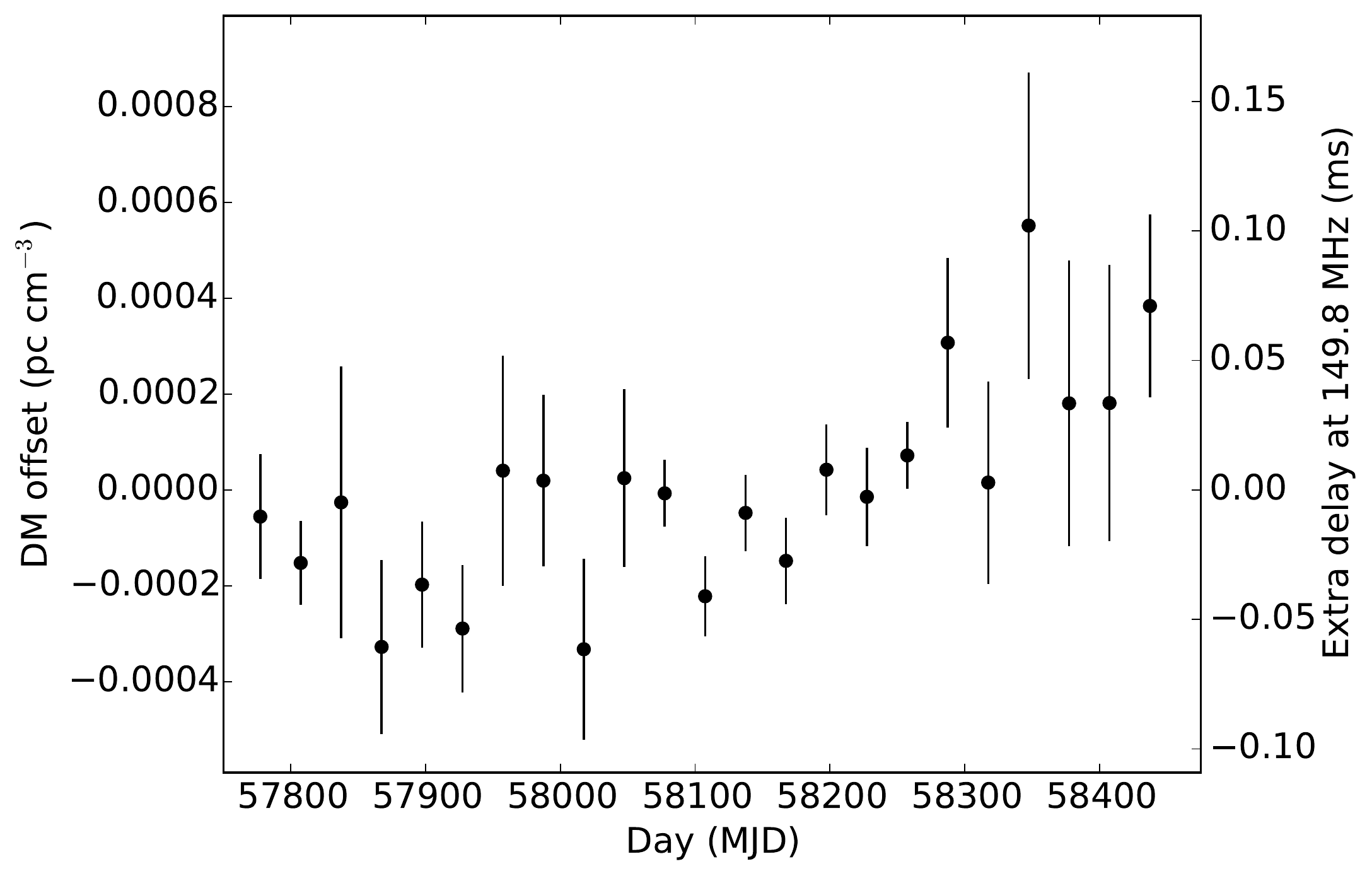}
\caption{The modelled DM variation of PSR J1658$+$3630 over the observing span.}
\label{fig:1658DMvariation}
\end{figure}

We modelled the DM variation of PSR\,J1658$+$3630 across the observing span by measuring the average DM value over spans of 30 days. The measured DM value in each epoch is then compared to the average across the observing span and is shown in Fig.~\ref{fig:1658DMvariation}. We found that for most of the observing span, the DM value fluctuates between $-0.0002$ and $+0.0001$ pc cm$^{-3}$ around the average DM value of 3.0439 pc cm$^{-3}$, with the observations since MJD 58280 showing larger DM increase of between $+0.0002$ and $+0.0005$ pc cm$^{-3}$ more than average. The timing solution took into consideration the measured DM variation. The other pulsars studied in this work are likely to show DM variation as well. However, due to their lower timing precision compared to PSR\,J1658$+$3630, the variations are too subtle to be measured.

PSR\,J1658+3630 is coincident (within $1.2\arcsec$) with SDSS\,J165826.50$+$363031.1, a $r=22.1$, $g-r=0.56$ star in the Sloan Digital Sky Survey~\citep[SDSS,][]{yaa+00,aaa+17}. This optical counterpart is also seen in images from the Panoramic Sky Survey Telescope and Rapid Response System~\citep[Pan-STARRS,][]{cmm+16,fmc+16}. Using white dwarf cooling models by~\citet{bwd+11}\footnote{\url{http://www.astro.umontreal.ca/~bergeron/CoolingModels/}}, as described in \citet{hb06,ks06,tbg11} and \citet{bwd+11}, we find that the SDSS $u-g$, $g-r$ and $r-i$ colors of the counterpart are consistent with those of a 0.9\,M$_\odot$, $T_\mathrm{eff}\sim5250$\,K white dwarf. Given this, and the low stellar density in SDSS (of order 9 stars per square arcminute with $r<23$), we conclude that the counterpart is the white dwarf companion of PSR J1658+3630. Given it's high mass of $M_\mathrm{WD}>0.87$\,M$_\odot$, it is likely to be a Carbon-Oxygen or a Oxygen-Neon-Magnesium type white dwarf. We found that the position of the pulsar obtained from the timing solution is different from both the SDSS observation taken in 2000 and the Pan-STARRS observation taken in 2014, suggesting that the system has moved across the sky over the years. With this assumption, we derived a proper motion of $+0\farcs015 \pm 0\farcs008$ yr$^{-1}$ in right ascension and $-0\farcs065 \pm 0\farcs008$ yr$^{-1}$ in declination. The overall proper motion indicates the pulsar has a transverse velocity of between 69-154 km s$^{-1}$ based on the DM-distances obtained. The proper motion estimated here is large enough that a sinusoidal variation with increasing amplitude will be seen in the timing data if it is not accounted for. As we only have one and a half years of timing data, we were not able to refine the measured proper motion through pulsar timing yet. Hence, we fixed the proper motion of the pulsar at the values measured while refining the other properties.

While the timing model is able to broadly describe the rotational and orbital properties of the pulsar, we note that there remains a significant timing residual in the TOAs and a poor $\chi_\mathrm{red}^2$ value. There are several possible contributions to the large timing residuals. First, the proper motion measured has a significant uncertainty of $0\farcs008$ yr$^{-1}$ in both right ascension and declination. The proper motion can be measured to a higher precision with pulsar timing, but this will require a longer observing span on the order of several years. We also noticed that there is an unmodelled extra delay on the timing residuals at around MJD 58165. This is likely a short duration increase in the DM of the pulsar on a time-scale shorter than the 30 days used to model the DM variation of the pulsar.

\begin{figure}
\centering
\includegraphics[width=\linewidth]{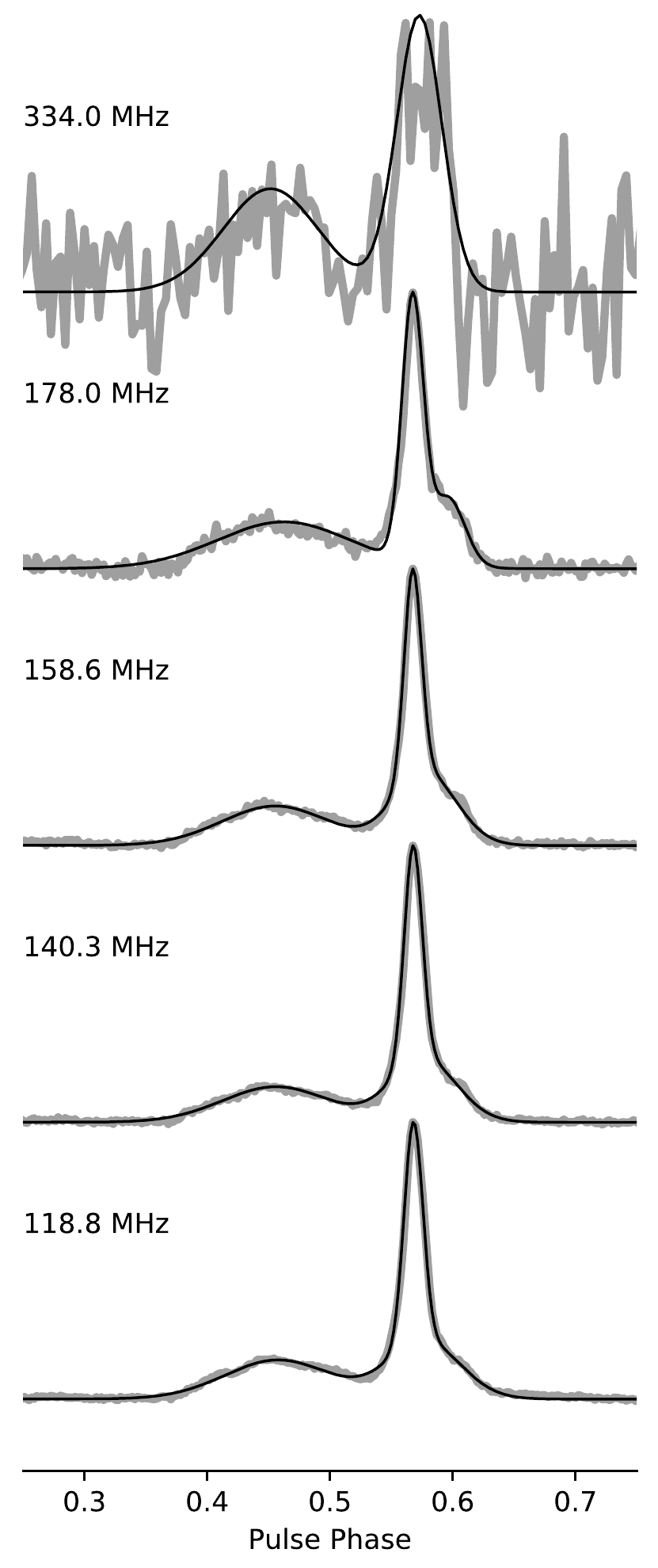}
\caption[The integrated pulse profiles of J1658$+$3630]{The integrated pulse profiles of PSR J1658$+$3630 at several different sub-bands obtained with the LOFAR core and at 334\,MHz obtained with the Lovell Telescope. The profiles (grey) are overlaid with the models (black) that describe the profiles at different observing frequencies in order to study profile evolution.}
\label{fig:J1658profiles}
\end{figure}

The average pulse profile integrated over the observing span shows frequency-dependent profile evolution. Fig.~\ref{fig:J1658profiles} shows the integrated pulse profiles at four different LOFAR sub-bands and at 334\,MHz, overlaid with the model that best describes the pulse profiles. The profiles in the LOFAR bandwidth are described by three von Mises components, corresponding to the leading bump, the main peak and a small trailing component that increases in relative intensity to the main peak as the observing frequency increases. The profile at 334\,MHz has a much lower S/N and was fitted with just 2 components. We are unable to identify if the trailing component is present at 334\,MHz. We also measured the width of the pulses at 50\% and 10\% of the maximum, shown in Table~\ref{tab:J1658fluxwidth}.

\begin{table*}
\centering
\caption[The flux densities and pulse widths of PSR\,J1658$+$3630 at different observing frequencies.]{The flux densities and pulse widths of the PSR\,J1658$+$3630 at 4 different LOFAR sub-bands of 118.8, 140.3, 158.6 and 178.0\,MHz and at 334\,MHz measured with the detection with the Lovell telescope. The upper limit of the flux density of PSR\,J1658$+$3630 based on the non-detection at 1532\,MHz is also quoted, estimated using the average W$_{50}$ value of the detection made at 334\,MHz, an 1-hour integration time and a detection S/N threshold of 10.}
\begin{tabular}{cccccc}
\hline
Frequency & Flux density & W$_{10}$ & $\delta_{10}$ & W$_{50}$ & $\delta_{50}$\\
MHz & mJy & ms & \% & ms & \%\\
\hline
118.8 & 24(8) & 6.282(16) & 19.03(5) & 0.6977(10)  & 2.113(3)\\
140.3 & 17(6) & 6.12(2) & 18.54(6) & 0.6768(11) & 2.050(3)\\
158.6 & 10(4) & 6.41(2) & 19.40(6) & 0.6712(15) & 2.033(5)\\
178.0 & 6(2) & 6.89(5) & 20.88(15) & 0.744(3) & 2.252(9)\\
334.0 & 1.7(3) & 7.42(19) & 22.4(6) & 1.5(2) & 4.4(7)\\
1532.0 & $<$0.06 & -- & -- & -- & --\\
\hline
\end{tabular}
\label{tab:J1658fluxwidth}
\end{table*}

The pulsar is also observed to undergo diffractive scintillation in some of the observations, where the scintillation bandwidth appeared to be smaller than the bandwidth of a single sub-band of 195\,kHz. This is much smaller than the expected bandwidth of 1.4\,MHz predicted by the NE2001 model. While the detailed study of the scintillation properties of the pulsar will be presented in a separate paper, we note that the presence of diffractive scintillation might affect the overall pulse profile shape of individual observations due to the variation in flux densities in different parts of the bandwidth at different observing epochs. This will subsequently affect the timing precision of these observations due to profile shape changes relative to the template used. The DM variation can affect the overall shape of the profile as well, as different parts of the bandwidth may not be aligned at the fiducial point of the template. 

\begin{figure}
\centering
\includegraphics[width=\linewidth]{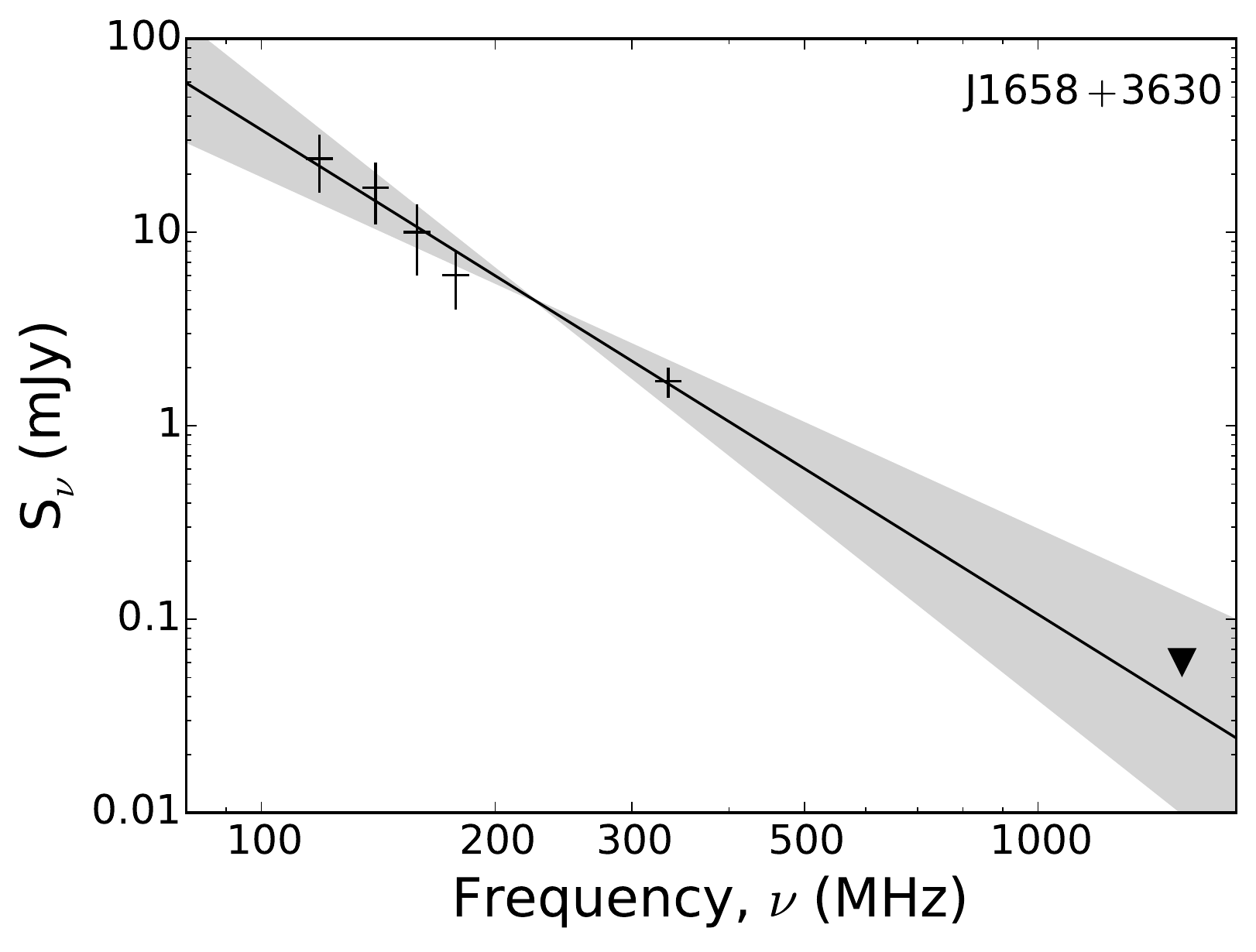}
\caption[The spectrum of PSR\,J1658$+$3630]{The spectrum of PSR\,J1658$+$3630, modelled with a single power law. The triangle is the upper limit in flux density at 1532\,MHz estimated from a 1-hour non-detection with the Lovell telescope. The black line is the best-fit spectral index of $-$2.5 and the grey shaded region is the 1$\sigma$ uncertainty in the fitted spectral index.}
\label{fig:J1658spectrum}
\end{figure}

We also calculated the average flux densities of PSR\,J1658$+$3630 at the same four LOFAR sub-bands and the flux density from the 334\,MHz observation in which the pulsar is detected. The average flux densities at the LOFAR frequencies were calculated from 41 observations obtained from the second dense campaign and are shown in Table~\ref{tab:J1658fluxwidth}. The diffractive scintillation and possibly refractive scintillation will result in variation in the flux densities. We model the spectral index of the pulsar using a single power law, shown in Fig.~\ref{fig:J1658spectrum}, and measured $\alpha = -2.5 \pm 0.7$. The pulsar is therefore a steep spectrum source, similar to the other MSPs discovered by LOFAR~\citep{bph+17,pbh+17}. The spectral index predicts a radio luminosity at 1400\,MHz, $L_{1400}$ of between 0.0015 and 0.007 mJy kpc$^2$, depending on the distance obtained by the electron density model used~\citep{cl02,ymw17}. This is similar to another recently discovered nearby, low-luminosity MSP J2322$-$2650~\citep[$L_{1400} = 0.008$ mJy kpc$^{-2}$;][]{sbb+18}, strengthening the argument that there could be a large population of low-luminosity Galactic MSPs that can only be detected at very low radio frequencies.

\section{Conclusions}

We have presented the properties of 21 pulsars discovered during the LOFAR Tied-Array All-Sky Survey (LOTAAS). We have obtained the rotational properties of the pulsars, which suggest that most of the pulsars discovered are part of an older population with $\tau_c \geq 10$\,Myr. We have also studied the frequency evolution of the integrated pulse profiles of the pulsars, where we found that PSR\,J1643$+$1338 showed frequency evolution that is contrary to the RFM model. Furthermore, we have measured the spectral indices of the pulsars and found that the average spectral index of the sample is lower than the known pulsar population, possibly linking the spectral properties with the ages of the pulsars. However, a larger sample of pulsars is required to confirm this relationship.

PSR\,J1657$+$3304 showed variation of a factor of 10 in the observed flux density over a span of 300 days. The pulsar also showed both nulling and mode changing on the order of minutes. We found that interstellar scintillation is unlikely to play a part in the long term flux density variation as the flux density variation is consistent across the fractionally wide LOFAR band.

Finally, we modelled the various properties of PSR\,J1658$+$3630, the first binary pulsar discovered by LOTAAS, showing that the pulsar is in a binary system with a companion with a minimum mass of 0.87$M_{\odot}$, likely a Carbon-Oxygen or Oxygen-Neon-Magnesium type white dwarf. The pulsar is also found to be an ultra steep spectrum source, with a spectral index of $\alpha=-2.5 \pm 0.7$, giving an estimated radio luminosity at 1400\,MHz between 0.0015 and 0.007 mJy kpc$^2$. This suggests that it could be part of a population of nearby, low-luminosity millisecond pulsars. 

\section*{Acknowledgements}

This paper is based in part on data obtained with the International LOFAR Telescope (ILT) under project codes LT5\_003 (PI: Verbiest), LC8\_035, LC9\_018, LC9\_021, LC9\_041 (PI: Verbiest) and LT10\_015. LOFAR~\citep{hwg+13} is the Low Frequency Array designed and constructed by ASTRON. It has observing, data processing, and data storage facilities in several countries, that are owned by various parties (each with their own funding sources), and that are collectively operated by the ILT foundation under a joint scientific policy. The ILT resources have benefited from the following recent major funding sources: CNRS-INSU, Observatoire de Paris and Universit\'{e} d'Orl\'{e}ans, France; BMBF, MIWF-NRW, MPG, Germany; Science Foundation Ireland (SFI), Department of Business, Enterprise and Innovation (DBEI), Ireland; NWO, The Netherlands; The Science and Technology Facilities Council, UK; Ministry of Science and Higher Education, Poland.
This paper is based in part on data obtained with the German stations of the International LOFAR Telescope (ILT), constructed by ASTRON~\citep{hwg+13}, during station-owners time. In this work we made use of data from the Unterweilenbach (DE602) LOFAR station funded by the Max-Planck-Institut f\"{u}r Astrophysik, Garching; the Tautenburg (DE603) LOFAR station funded by the State of Thuringia, supported by the European Union (EFRE) and the Federal Ministry of Education and Research (BMBF) Verbundforschung project D-LOFAR I (grant 05A08ST1); the Potsdam (DE604) LOFAR station funded by the Leibniz-Institut f\"{u}r Astrophysik, Potsdam; the J\"{u}lich (DE605) LOFAR station supported by the BMBF Verbundforschung project D-LOFAR I (grant 05A08LJ1); and the Norderstedt (DE609) LOFAR station funded by the BMBF Verbundforschung project D-LOFAR II (grant 05A11LJ1). The observations of the German LOFAR stations were carried out in the stand-alone GLOW mode (German LOng-Wavelength array), which is technically operated and supported by the Max-Planck-Institut f\"{u}r Radioastronomie, the Forschungszentrum J\"{u}lich and Bielefeld University. We acknowledge support and operation of the GLOW network, computing and storage facilities by the FZ-J\"{u}lich, the MPIfR and Bielefeld University and financial support from BMBF D-LOFAR III (grant 05A14PBA), and by the states of Nordrhein-Westfalia and Hamburg.
JWTH., DM, VIK., CGB and SS acknowledge funding from an NWO Vidi fellowship and from the European Research Council under the European Union's Seventh Framework Programme (FP/2007-2013) / ERC Starting Grant agreement nr. 337062 (`DRAGNET'). DM is a Banting fellow. JvL acknowledges funding from the European Research Council under the European Union’s Seventh Framework Programme (FP/2007-2013) / ERC Grant Agreement n. 617199 (`ALERT'), and from Vici research programme `ARGO' with project number 639.043.815, financed by the Netherlands Organisation for Scientific Research (NWO). 




\bibliographystyle{mnras}
\bibliography{references} 

\begin{thebibliography}{}
\makeatletter
\relax
\def\mn@urlcharsother{\let\do\@makeother \do\$\do\&\do\#\do\^\do\_\do\%\do\~}
\def\mn@doi{\begingroup\mn@urlcharsother \@ifnextchar [ {\mn@doi@}
  {\mn@doi@[]}}
\def\mn@doi@[#1]#2{\def\@tempa{#1}\ifx\@tempa\@empty \href
  {http://dx.doi.org/#2} {doi:#2}\else \href {http://dx.doi.org/#2} {#1}\fi
  \endgroup}
\def\mn@eprint#1#2{\mn@eprint@#1:#2::\@nil}
\def\mn@eprint@arXiv#1{\href {http://arxiv.org/abs/#1} {{\tt arXiv:#1}}}
\def\mn@eprint@dblp#1{\href {http://dblp.uni-trier.de/rec/bibtex/#1.xml}
  {dblp:#1}}
\def\mn@eprint@#1:#2:#3:#4\@nil{\def\@tempa {#1}\def\@tempb {#2}\def\@tempc
  {#3}\ifx \@tempc \@empty \let \@tempc \@tempb \let \@tempb \@tempa \fi \ifx
  \@tempb \@empty \def\@tempb {arXiv}\fi \@ifundefined
  {mn@eprint@\@tempb}{\@tempb:\@tempc}{\expandafter \expandafter \csname
  mn@eprint@\@tempb\endcsname \expandafter{\@tempc}}}

\bibitem[\protect\citeauthoryear{{Albareti} et~al.,}{{Albareti}
  et~al.}{2017}]{aaa+17}
{Albareti} F.~D.,  et~al., 2017, \mn@doi [\apjs] {10.3847/1538-4365/aa8992},
  \href {http://adsabs.harvard.edu/abs/2017ApJS..233...25A} {233, 25}

\bibitem[\protect\citeauthoryear{{Bassa} et~al.,}{{Bassa}
  et~al.}{2017}]{bph+17}
{Bassa} C.~G.,  et~al., 2017, \mn@doi [\apj] {10.3847/2041-8213/aa8400}, \href
  {https://ui.adsabs.harvard.edu/#abs/2017ApJ...846L..20B} {846, L20}

\bibitem[\protect\citeauthoryear{{Bates}, {Lorimer}  \& {Verbiest}}{{Bates}
  et~al.}{2013}]{blv13}
{Bates} S.~D.,  {Lorimer} D.~R.,   {Verbiest} J.~P.~W.,  2013, \mn@doi [\mnras]
  {10.1093/mnras/stt257}, \href
  {http://adsabs.harvard.edu/abs/2013MNRAS.431.1352B} {431, 1352}

\bibitem[\protect\citeauthoryear{{Bergeron} et~al.,}{{Bergeron}
  et~al.}{2011}]{bwd+11}
{Bergeron} P.,  et~al., 2011, \mn@doi [\apj] {10.1088/0004-637X/737/1/28},
  \href {http://adsabs.harvard.edu/abs/2011ApJ...737...28B} {737, 28}

\bibitem[\protect\citeauthoryear{{Bilous} et~al.,}{{Bilous}
  et~al.}{2016}]{bkk+16}
{Bilous} A.~V.,  et~al., 2016, \mn@doi [\aap] {10.1051/0004-6361/201527702},
  \href {http://adsabs.harvard.edu/abs/2016A%26A...591A.134B} {591, A134}

\bibitem[\protect\citeauthoryear{{Chambers} et~al.,}{{Chambers}
  et~al.}{2016}]{cmm+16}
{Chambers} K.~C.,  et~al., 2016, preprint, \href
  {http://adsabs.harvard.edu/abs/2016arXiv161205560C} {} (\mn@eprint {arXiv}
  {1612.05560})

\bibitem[\protect\citeauthoryear{{Chen} \& {Ruderman}}{{Chen} \&
  {Ruderman}}{1993}]{cr93}
{Chen} K.,  {Ruderman} M.,  1993, \mn@doi [\apj] {10.1086/172129}, \href
  {http://adsabs.harvard.edu/abs/1993ApJ...402..264C} {402, 264}

\bibitem[\protect\citeauthoryear{{Chen} \& {Wang}}{{Chen} \&
  {Wang}}{2014}]{cw14}
{Chen} J.~L.,  {Wang} H.~G.,  2014, \mn@doi [\apjs]
  {10.1088/0067-0049/215/1/11}, \href
  {https://ui.adsabs.harvard.edu/abs/2014ApJS..215...11C} {215, 11}

\bibitem[\protect\citeauthoryear{{Coenen} et~al.,}{{Coenen}
  et~al.}{2014}]{clh+14}
{Coenen} T.,  et~al., 2014, \mn@doi [\aap] {10.1051/0004-6361/201424495}, \href
  {http://adsabs.harvard.edu/abs/2014A%26A...570A..60C} {570, A60}

\bibitem[\protect\citeauthoryear{{Cordes}}{{Cordes}}{1978}]{cor78}
{Cordes} J.~M.,  1978, \mn@doi [\apj] {10.1086/156218}, \href
  {https://ui.adsabs.harvard.edu/abs/1978ApJ...222.1006C} {222, 1006}

\bibitem[\protect\citeauthoryear{{Cordes} \& {Lazio}}{{Cordes} \&
  {Lazio}}{2002}]{cl02}
{Cordes} J.~M.,  {Lazio} T.~J.~W.,  2002, preprint, \href
  {http://adsabs.harvard.edu/abs/2002astro.ph..7156C} {} (\mn@eprint {arXiv}
  {0207156})

\bibitem[\protect\citeauthoryear{{Cronyn}}{{Cronyn}}{1970}]{cro70}
{Cronyn} W.~M.,  1970, \mn@doi [Science] {10.1126/science.168.3938.1453}, \href
  {https://ui.adsabs.harvard.edu/abs/1970Sci...168.1453C} {168, 1453}

\bibitem[\protect\citeauthoryear{{Demorest}, {Pennucci}, {Ransom}, {Roberts}
  \& {Hessels}}{{Demorest} et~al.}{2010}]{dpr+10}
{Demorest} P.~B.,  {Pennucci} T.,  {Ransom} S.~M.,  {Roberts} M.~S.~E.,
  {Hessels} J.~W.~T.,  2010, \mn@doi [\nat] {10.1038/nature09466}, \href
  {http://adsabs.harvard.edu/abs/2010Natur.467.1081D} {467, 1081}

\bibitem[\protect\citeauthoryear{{Edwards}, {Hobbs}  \& {Manchester}}{{Edwards}
  et~al.}{2006}]{ehm06}
{Edwards} R.~T.,  {Hobbs} G.~B.,   {Manchester} R.~N.,  2006, \mn@doi [\mnras]
  {10.1111/j.1365-2966.2006.10870.x}, \href
  {http://adsabs.harvard.edu/abs/2006MNRAS.372.1549E} {372, 1549}

\bibitem[\protect\citeauthoryear{{Evans}, {Hastings}  \& {Peacock}}{{Evans}
  et~al.}{2000}]{ehp00}
{Evans} M.,  {Hastings} N.,   {Peacock} B.,  2000, {Statistical Distributions},
  3rd edn.
Wiley, New York

\bibitem[\protect\citeauthoryear{{Flewelling} et~al.,}{{Flewelling}
  et~al.}{2016}]{fmc+16}
{Flewelling} H.~A.,  et~al., 2016, arXiv e-prints, \href
  {https://ui.adsabs.harvard.edu/\#abs/2016arXiv161205243F} {p.
  arXiv:1612.05243}

\bibitem[\protect\citeauthoryear{{Geyer} \& {Karastergiou}}{{Geyer} \&
  {Karastergiou}}{2016}]{gk16}
{Geyer} M.,  {Karastergiou} A.,  2016, \mn@doi [\mnras]
  {10.1093/mnras/stw1724}, \href
  {https://ui.adsabs.harvard.edu/abs/2016MNRAS.462.2587G} {462, 2587}

\bibitem[\protect\citeauthoryear{{Geyer} et~al.,}{{Geyer}
  et~al.}{2017}]{gkk+17}
{Geyer} M.,  et~al., 2017, \mn@doi [\mnras] {10.1093/mnras/stx1151}, \href
  {http://adsabs.harvard.edu/abs/2017MNRAS.470.2659G} {470, 2659}

\bibitem[\protect\citeauthoryear{{Haslam}, {Salter}, {Stoffel}  \&
  {Wilson}}{{Haslam} et~al.}{1982}]{hssw82}
{Haslam} C.~G.~T.,  {Salter} C.~J.,  {Stoffel} H.,   {Wilson} W.~E.,  1982,
  \aaps, \href {http://adsabs.harvard.edu/abs/1982A%26AS...47....1H} {47, 1}

\bibitem[\protect\citeauthoryear{{Hassall} et~al.,}{{Hassall}
  et~al.}{2012}]{hsh+12}
{Hassall} T.~E.,  et~al., 2012, \mn@doi [\aap] {10.1051/0004-6361/201218970},
  \href {http://adsabs.harvard.edu/abs/2012A%26A...543A..66H} {543, A66}

\bibitem[\protect\citeauthoryear{{Hewish}, {Bell}, {Pilkington}, {Scott}  \&
  {Collins}}{{Hewish} et~al.}{1968}]{hbp+68}
{Hewish} A.,  {Bell} S.~J.,  {Pilkington} J.~D.~H.,  {Scott} P.~F.,   {Collins}
  R.~A.,  1968, \mn@doi [\nat] {10.1038/217709a0}, \href
  {https://ui.adsabs.harvard.edu/abs/1968Natur.217..709H} {217, 709}

\bibitem[\protect\citeauthoryear{{Hobbs}, {Edwards}  \& {Manchester}}{{Hobbs}
  et~al.}{2006}]{hem06}
{Hobbs} G.~B.,  {Edwards} R.~T.,   {Manchester} R.~N.,  2006, \mn@doi [\mnras]
  {10.1111/j.1365-2966.2006.10302.x}, \href
  {http://adsabs.harvard.edu/abs/2006MNRAS.369..655H} {369, 655}

\bibitem[\protect\citeauthoryear{{Hobbs}, {Lyne}  \& {Kramer}}{{Hobbs}
  et~al.}{2010}]{hlk10}
{Hobbs} G.,  {Lyne} A.~G.,   {Kramer} M.,  2010, \mn@doi [\mnras]
  {10.1111/j.1365-2966.2009.15938.x}, \href
  {https://ui.adsabs.harvard.edu/abs/2010MNRAS.402.1027H} {402, 1027}

\bibitem[\protect\citeauthoryear{{Holberg} \& {Bergeron}}{{Holberg} \&
  {Bergeron}}{2006}]{hb06}
{Holberg} J.~B.,  {Bergeron} P.,  2006, \mn@doi [\aj] {10.1086/505938}, \href
  {http://adsabs.harvard.edu/abs/2006AJ....132.1221H} {132, 1221}

\bibitem[\protect\citeauthoryear{{Hotan}, {van Straten}  \&
  {Manchester}}{{Hotan} et~al.}{2004}]{hsm04}
{Hotan} A.~W.,  {van Straten} W.,   {Manchester} R.~N.,  2004, \mn@doi [\pasa]
  {10.1071/AS04022}, \href {http://adsabs.harvard.edu/abs/2004PASA...21..302H}
  {21, 302}

\bibitem[\protect\citeauthoryear{{Jankowski}, {van Straten}, {Keane}, {Bailes},
  {Barr}, {Johnston}  \& {Kerr}}{{Jankowski} et~al.}{2018}]{jsk+18}
{Jankowski} F.,  {van Straten} W.,  {Keane} E.~F.,  {Bailes} M.,  {Barr} E.~D.,
   {Johnston} S.,   {Kerr} M.,  2018, \mn@doi [\mnras] {10.1093/mnras/stx2476},
  \href {https://ui.adsabs.harvard.edu/\#abs/2018MNRAS.473.4436J} {473, 4436}

\bibitem[\protect\citeauthoryear{{Kijak}, {Lewandowski}, {Maron}, {Gupta}  \&
  {Jessner}}{{Kijak} et~al.}{2011}]{klm+11}
{Kijak} J.,  {Lewandowski} W.,  {Maron} O.,  {Gupta} Y.,   {Jessner} A.,  2011,
  \mn@doi [\aap] {10.1051/0004-6361/201014274}, \href
  {https://ui.adsabs.harvard.edu/\#abs/2011A&A...531A..16K} {531, A16}

\bibitem[\protect\citeauthoryear{{Kondratiev} et~al.,}{{Kondratiev}
  et~al.}{2016}]{kvh+16}
{Kondratiev} V.~I.,  et~al., 2016, \mn@doi [\aap]
  {10.1051/0004-6361/201527178}, \href
  {http://adsabs.harvard.edu/abs/2016A%26A...585A.128K} {585, A128}

\bibitem[\protect\citeauthoryear{{Kowalski} \& {Saumon}}{{Kowalski} \&
  {Saumon}}{2006}]{ks06}
{Kowalski} P.~M.,  {Saumon} D.,  2006, \mn@doi [\apjl] {10.1086/509723}, \href
  {http://adsabs.harvard.edu/abs/2006ApJ...651L.137K} {651, L137}

\bibitem[\protect\citeauthoryear{{Kramer}, {Wielebinski}, {Jessner}, {Gil}  \&
  {Seiradakis}}{{Kramer} et~al.}{1994}]{kwj+94}
{Kramer} M.,  {Wielebinski} R.,  {Jessner} A.,  {Gil} J.~A.,   {Seiradakis}
  J.~H.,  1994, \aaps, \href
  {https://ui.adsabs.harvard.edu/abs/1994A&AS..107..515K} {107, 515}

\bibitem[\protect\citeauthoryear{{Kramer}, {Lyne}, {O'Brien}, {Jordan}  \&
  {Lorimer}}{{Kramer} et~al.}{2006}]{klb+06}
{Kramer} M.,  {Lyne} A.~G.,  {O'Brien} J.~T.,  {Jordan} C.~A.,   {Lorimer}
  D.~R.,  2006, \mn@doi [Science] {10.1126/science.1124060}, \href
  {https://ui.adsabs.harvard.edu/abs/2006Sci...312..549K} {312, 549}

\bibitem[\protect\citeauthoryear{{Lang}}{{Lang}}{1971}]{lan71}
{Lang} K.~R.,  1971, \mn@doi [\apj] {10.1086/150836}, \href
  {https://ui.adsabs.harvard.edu/abs/1971ApJ...164..249L} {164, 249}

\bibitem[\protect\citeauthoryear{{Lange}, {Camilo}, {Wex}, {Kramer}, {Backer},
  {Lyne}  \& {Doroshenko}}{{Lange} et~al.}{2001}]{lcw+01}
{Lange} C.,  {Camilo} F.,  {Wex} N.,  {Kramer} M.,  {Backer} D.~C.,  {Lyne}
  A.~G.,   {Doroshenko} O.,  2001, \mnras, \href
  {http://cdsads.u-strasbg.fr/cgi-bin/nph-bib_query?bibcode=2001MNRAS
  .326..274L&db_key=AST} {326, 274}

\bibitem[\protect\citeauthoryear{{Lawson}, {Mayer}, {Osborne}  \&
  {Parkinson}}{{Lawson} et~al.}{1987}]{lmop87}
{Lawson} K.~D.,  {Mayer} C.~J.,  {Osborne} J.~L.,   {Parkinson} M.~L.,  1987,
  \mn@doi [\mnras] {10.1093/mnras/225.2.307}, \href
  {https://ui.adsabs.harvard.edu/#abs/1987MNRAS.225..307L} {225, 307}

\bibitem[\protect\citeauthoryear{{Lee} \& {Jokipii}}{{Lee} \&
  {Jokipii}}{1976}]{lj76}
{Lee} L.~C.,  {Jokipii} J.~R.,  1976, \mn@doi [\apj] {10.1086/154434}, \href
  {https://ui.adsabs.harvard.edu/abs/1976ApJ...206..735L} {206, 735}

\bibitem[\protect\citeauthoryear{{L{\"o}hmer}, {Mitra}, {Gupta}, {Kramer}  \&
  {Ahuja}}{{L{\"o}hmer} et~al.}{2004}]{lmg+04}
{L{\"o}hmer} O.,  {Mitra} D.,  {Gupta} Y.,  {Kramer} M.,   {Ahuja} A.,  2004,
  \mn@doi [\aap] {10.1051/0004-6361:20035881}, \href
  {https://ui.adsabs.harvard.edu/abs/2004A&A...425..569L} {425, 569}

\bibitem[\protect\citeauthoryear{{Lorimer} \& {Kramer}}{{Lorimer} \&
  {Kramer}}{2005}]{lk05}
{Lorimer} D.~R.,  {Kramer} M.,  2005, {Handbook of Pulsar Astronomy}

\bibitem[\protect\citeauthoryear{{Lyne}, {Hobbs}, {Kramer}, {Stairs}  \&
  {Stappers}}{{Lyne} et~al.}{2010}]{lhk+10}
{Lyne} A.,  {Hobbs} G.,  {Kramer} M.,  {Stairs} I.,   {Stappers} B.,  2010,
  \mn@doi [Science] {10.1126/science.1186683}, \href
  {https://ui.adsabs.harvard.edu/abs/2010Sci...329..408L} {329, 408}

\bibitem[\protect\citeauthoryear{{Maron}, {Kijak}, {Kramer}  \&
  {Wielebinski}}{{Maron} et~al.}{2000}]{mkkw00}
{Maron} O.,  {Kijak} J.,  {Kramer} M.,   {Wielebinski} R.,  2000, \mn@doi
  [\aaps] {10.1051/aas:2000298}, \href
  {http://adsabs.harvard.edu/abs/2000A%26AS..147..195M} {147, 195}

\bibitem[\protect\citeauthoryear{{McLaughlin} et~al.,}{{McLaughlin}
  et~al.}{2006}]{mll+06}
{McLaughlin} M.~A.,  et~al., 2006, \mn@doi [\nat] {10.1038/nature04440}, \href
  {http://adsabs.harvard.edu/abs/2006Natur.439..817M} {439, 817}

\bibitem[\protect\citeauthoryear{{Michilli} et~al.,}{{Michilli}
  et~al.}{2019}]{mbc+19}
{Michilli} D.,  et~al., 2019, \mn@doi [\mnras] {10.1093/mnras/stz2997}, \href
  {https://ui.adsabs.harvard.edu/abs/2019MNRAS.tmp.2570M} {p.~2570}

\bibitem[\protect\citeauthoryear{{Noutsos} et~al.,}{{Noutsos}
  et~al.}{2015}]{nsk+15}
{Noutsos} A.,  et~al., 2015, \mn@doi [\aap] {10.1051/0004-6361/201425186},
  \href {https://ui.adsabs.harvard.edu/abs/2015A&A...576A..62N} {576, A62}

\bibitem[\protect\citeauthoryear{{Perera}, {Stappers}, {Weltevrede}, {Lyne}  \&
  {Bassa}}{{Perera} et~al.}{2015}]{psw+15}
{Perera} B.~B.~P.,  {Stappers} B.~W.,  {Weltevrede} P.,  {Lyne} A.~G.,
  {Bassa} C.~G.,  2015, \mn@doi [\mnras] {10.1093/mnras/stu2187}, \href
  {https://ui.adsabs.harvard.edu/abs/2015MNRAS.446.1380P} {446, 1380}

\bibitem[\protect\citeauthoryear{{Pilia} et~al.,}{{Pilia}
  et~al.}{2016}]{phs+16}
{Pilia} M.,  et~al., 2016, \mn@doi [\aap] {10.1051/0004-6361/201425196}, \href
  {https://ui.adsabs.harvard.edu/#abs/2016A&A...586A..92P} {586, A92}

\bibitem[\protect\citeauthoryear{{Pleunis} et~al.,}{{Pleunis}
  et~al.}{2017}]{pbh+17}
{Pleunis} Z.,  et~al., 2017, \mn@doi [\apj] {10.1051/0004-6361/201628536},
  \href {http://adsabs.harvard.edu/abs/2017A%26A...598A..78I} {700, 1}

\bibitem[\protect\citeauthoryear{{Rajwade}, {Lorimer}  \& {Anderson}}{{Rajwade}
  et~al.}{2016}]{rla16}
{Rajwade} K.,  {Lorimer} D.~R.,   {Anderson} L.~D.,  2016, \mn@doi [\mnras]
  {10.1093/mnras/stv2334}, \href
  {https://ui.adsabs.harvard.edu/\#abs/2016MNRAS.455..493R} {455, 493}

\bibitem[\protect\citeauthoryear{{Rankin}, {Rodriguez}  \& {Wright}}{{Rankin}
  et~al.}{2006}]{rrw06}
{Rankin} J.~M.,  {Rodriguez} C.,   {Wright} G. A.~E.,  2006, \mn@doi [\mnras]
  {10.1111/j.1365-2966.2006.10512.x}, \href
  {https://ui.adsabs.harvard.edu/abs/2006MNRAS.370..673R} {370, 673}

\bibitem[\protect\citeauthoryear{{Reich} \& {Reich}}{{Reich} \&
  {Reich}}{1988}]{rr88}
{Reich} P.,  {Reich} W.,  1988, \aap, \href
  {https://ui.adsabs.harvard.edu/#abs/1988A&A...196..211R} {196, 211}

\bibitem[\protect\citeauthoryear{{Rickett}}{{Rickett}}{1977}]{ric77}
{Rickett} B.~J.,  1977, \mn@doi [\araa] {10.1146/annurev.aa.15.090177.002403},
  \href {https://ui.adsabs.harvard.edu/abs/1977ARA&A..15..479R} {15, 479}

\bibitem[\protect\citeauthoryear{{Rickett}, {Coles}  \& {Bourgois}}{{Rickett}
  et~al.}{1984}]{rcb84}
{Rickett} B.~J.,  {Coles} W.~A.,   {Bourgois} G.,  1984, \aap, \href
  {https://ui.adsabs.harvard.edu/abs/1984A&A...134..390R} {134, 390}

\bibitem[\protect\citeauthoryear{{Ruderman} \& {Sutherland}}{{Ruderman} \&
  {Sutherland}}{1975}]{rs75}
{Ruderman} M.~A.,  {Sutherland} P.~G.,  1975, \mn@doi [\apj] {10.1086/153393},
  \href {http://adsabs.harvard.edu/abs/1975ApJ...196...51R} {196, 51}

\bibitem[\protect\citeauthoryear{{Sanidas} et~al.,}{{Sanidas}
  et~al.}{2019}]{scb+19}
{Sanidas} S.,  et~al., 2019, arXiv e-prints, \href
  {https://ui.adsabs.harvard.edu/abs/2019arXiv190504977S} {p. arXiv:1905.04977}

\bibitem[\protect\citeauthoryear{{Sieber}}{{Sieber}}{1973}]{sie73}
{Sieber} W.,  1973, \aap, \href
  {https://ui.adsabs.harvard.edu/abs/1973A&A....28..237S} {28, 237}

\bibitem[\protect\citeauthoryear{{Spiewak} et~al.,}{{Spiewak}
  et~al.}{2018}]{sbb+18}
{Spiewak} R.,  et~al., 2018, \mn@doi [\mnras] {10.1093/mnras/stx3157}, \href
  {https://ui.adsabs.harvard.edu/abs/2018MNRAS.475..469S} {475, 469}

\bibitem[\protect\citeauthoryear{{Stappers} et~al.,}{{Stappers}
  et~al.}{2011}]{sha+11}
{Stappers} B.~W.,  et~al., 2011, \mn@doi [\aap] {10.1051/0004-6361/201116681},
  \href {http://adsabs.harvard.edu/abs/2011A%26A...530A..80S} {530, A80}

\bibitem[\protect\citeauthoryear{{Taylor} et~al.,}{{Taylor}
  et~al.}{2012}]{tek+12}
{Taylor} G.~B.,  et~al., 2012, \mn@doi [Journal of Astronomical
  Instrumentation] {10.1142/S2251171712500043}, \href
  {https://ui.adsabs.harvard.edu/abs/2012JAI.....150004T} {1, 1250004}

\bibitem[\protect\citeauthoryear{{Tingay} et~al.,}{{Tingay}
  et~al.}{2013}]{tgb+13}
{Tingay} S.~J.,  et~al., 2013, \mn@doi [\pasa] {10.1017/pasa.2012.007}, \href
  {http://adsabs.harvard.edu/abs/2013PASA...30....7T} {30, e007}

\bibitem[\protect\citeauthoryear{{Tremblay}, {Bergeron}  \&
  {Gianninas}}{{Tremblay} et~al.}{2011}]{tbg11}
{Tremblay} P.-E.,  {Bergeron} P.,   {Gianninas} A.,  2011, \mn@doi [\apj]
  {10.1088/0004-637X/730/2/128}, \href
  {http://adsabs.harvard.edu/abs/2011ApJ...730..128T} {730, 128}

\bibitem[\protect\citeauthoryear{{Tyul'bashev}, {Tyul'bashev}, {Oreshko}  \&
  {Logvinenko}}{{Tyul'bashev} et~al.}{2016}]{ttol16}
{Tyul'bashev} S.~A.,  {Tyul'bashev} V.~S.,  {Oreshko} V.~V.,   {Logvinenko}
  S.~V.,  2016, \mn@doi [Astronomy Reports] {10.1134/S1063772916020128}, \href
  {https://ui.adsabs.harvard.edu/\#abs/2016ARep...60..220T} {60, 220}

\bibitem[\protect\citeauthoryear{{Tyul'bashev} et~al.,}{{Tyul'bashev}
  et~al.}{2017}]{ttk+17}
{Tyul'bashev} S.~A.,  et~al., 2017, \mn@doi [Astronomy Reports]
  {10.1134/S1063772917100109}, \href
  {https://ui.adsabs.harvard.edu/\#abs/2017ARep...61..848T} {61, 848}

\bibitem[\protect\citeauthoryear{{Tyul'bashev}, {Tyul'bashev}  \&
  {Malofeev}}{{Tyul'bashev} et~al.}{2018}]{ttm18}
{Tyul'bashev} S.~A.,  {Tyul'bashev} V.~S.,   {Malofeev} V.~M.,  2018, \mn@doi
  [\aap] {10.1051/0004-6361/201833102}, \href
  {https://ui.adsabs.harvard.edu/\#abs/2018A&A...618A..70T} {618, A70}

\bibitem[\protect\citeauthoryear{{Wang}, {Manchester}  \& {Johnston}}{{Wang}
  et~al.}{2007}]{wmj07}
{Wang} N.,  {Manchester} R.~N.,   {Johnston} S.,  2007, \mn@doi [\mnras]
  {10.1111/j.1365-2966.2007.11703.x}, \href
  {https://ui.adsabs.harvard.edu/\#abs/2007MNRAS.377.1383W} {377, 1383}

\bibitem[\protect\citeauthoryear{{Yao}, {Manchester}  \& {Wang}}{{Yao}
  et~al.}{2017}]{ymw17}
{Yao} J.~M.,  {Manchester} R.~N.,   {Wang} N.,  2017, \mn@doi [\apj]
  {10.3847/1538-4357/835/1/29}, \href
  {http://adsabs.harvard.edu/abs/2017ApJ...835...29Y} {835, 29}

\bibitem[\protect\citeauthoryear{{York} et~al.,}{{York} et~al.}{2000}]{yaa+00}
{York} D.~G.,  et~al., 2000, \mn@doi [\aj] {10.1086/301513}, \href
  {http://adsabs.harvard.edu/abs/2000AJ....120.1579Y} {120, 1579}

\bibitem[\protect\citeauthoryear{{van Haarlem} et~al.,}{{van Haarlem}
  et~al.}{2013}]{hwg+13}
{van Haarlem} M.~P.,  et~al., 2013, \mn@doi [\aap]
  {10.1051/0004-6361/201220873}, \href
  {http://adsabs.harvard.edu/abs/2013A%26A...556A...2V} {556, A2}

\bibitem[\protect\citeauthoryear{{van Leeuwen} \& {Stappers}}{{van Leeuwen} \&
  {Stappers}}{2010}]{ls10}
{van Leeuwen} J.,  {Stappers} B.~W.,  2010, \mn@doi [\aap]
  {10.1051/0004-6361/200913121}, \href
  {http://adsabs.harvard.edu/abs/2010A%26A...509A...7V} {509, A7}

\bibitem[\protect\citeauthoryear{{van Straten} \& {Bailes}}{{van Straten} \&
  {Bailes}}{2011}]{sb10}
{van Straten} W.,  {Bailes} M.,  2011, \mn@doi [\pasa] {10.1071/AS10021}, \href
  {http://adsabs.harvard.edu/abs/2011PASA...28....1V} {28, 1}

\makeatother
\end{thebibliography}








\bsp	
\label{lastpage}
\end{document}